\documentclass[parskip=no]{article}

\usepackage{fzjdoc}

\RequirePackage{helvet}
\RequirePackage[T1]{fontenc}
\RequirePackage[utf8]{inputenc}
\RequirePackage[ngerman,english]{babel}
\RequirePackage{csquotes}
\RequirePackage{xcolor}
\RequirePackage{amsmath} 
\RequirePackage{amsfonts} 
\RequirePackage{amssymb}
\RequirePackage{graphicx}
\RequirePackage{longtable}
\RequirePackage[automark, headsepline]{scrlayer-scrpage}
\RequirePackage{tikz}
\RequirePackage{acronym}
\RequirePackage{setspace}
\RequirePackage{pdfpages}
\RequirePackage{lastpage}
\RequirePackage[hidelinks=false]{hyperref}
\RequirePackage{tabularx}
\RequirePackage{array,ragged2e}
\RequirePackage{ifthen}
\RequirePackage[absolute]{textpos}
\RequirePackage{translations}

\usepackage{cases}
\usepackage{float}
\usepackage{hyphenat}
\usepackage{mathtools}
\begin{document}

\newpage
\begin{center}
    \fontsize{18pt}{18pt}\selectfont
    \textbf{Extending the Flory-Huggins Theory for Crystalline Multicomponent Mixtures}
    \par
    \vspace{0.3cm}
    \fontsize{12pt}{12pt}\selectfont
    {Maxime Siber,$^{\ast}$\textit{$^{a,b}$} Olivier J. J. Ronsin,\textit{$^{a}$} and Jens Harting \textit{$^{a,b,c}$}}
\end{center}

\fontsize{9pt}{9pt}\selectfont
{\textit{$^{a}$~Helmholtz Institute Erlangen-Nürnberg for Renewable Energy, Forschungszentrum Jülich, Fürther Straße 248, 90429 Nürnberg, Germany, E-mail: m.siber@fz-juelich.de}}\par
{\textit{$^{b}$~Department of Chemical and Biological Engineering, Friedrich-Alexander-Universität Erlangen-Nürnberg, Fürther Straße 248, 90429 Nürnberg, Germany}}\par
{\textit{$^{c}$~Department of Physics, Friedrich-Alexander-Universität Erlangen-Nürnberg, Fürther Straße 248, 90429 Nürnberg, Germany}}

\fontsize{10pt}{10pt}\selectfont

\addtocounter{figure}{-1}
\begin{figure*}[h!]
    \centering
    \includegraphics[scale=0.25]{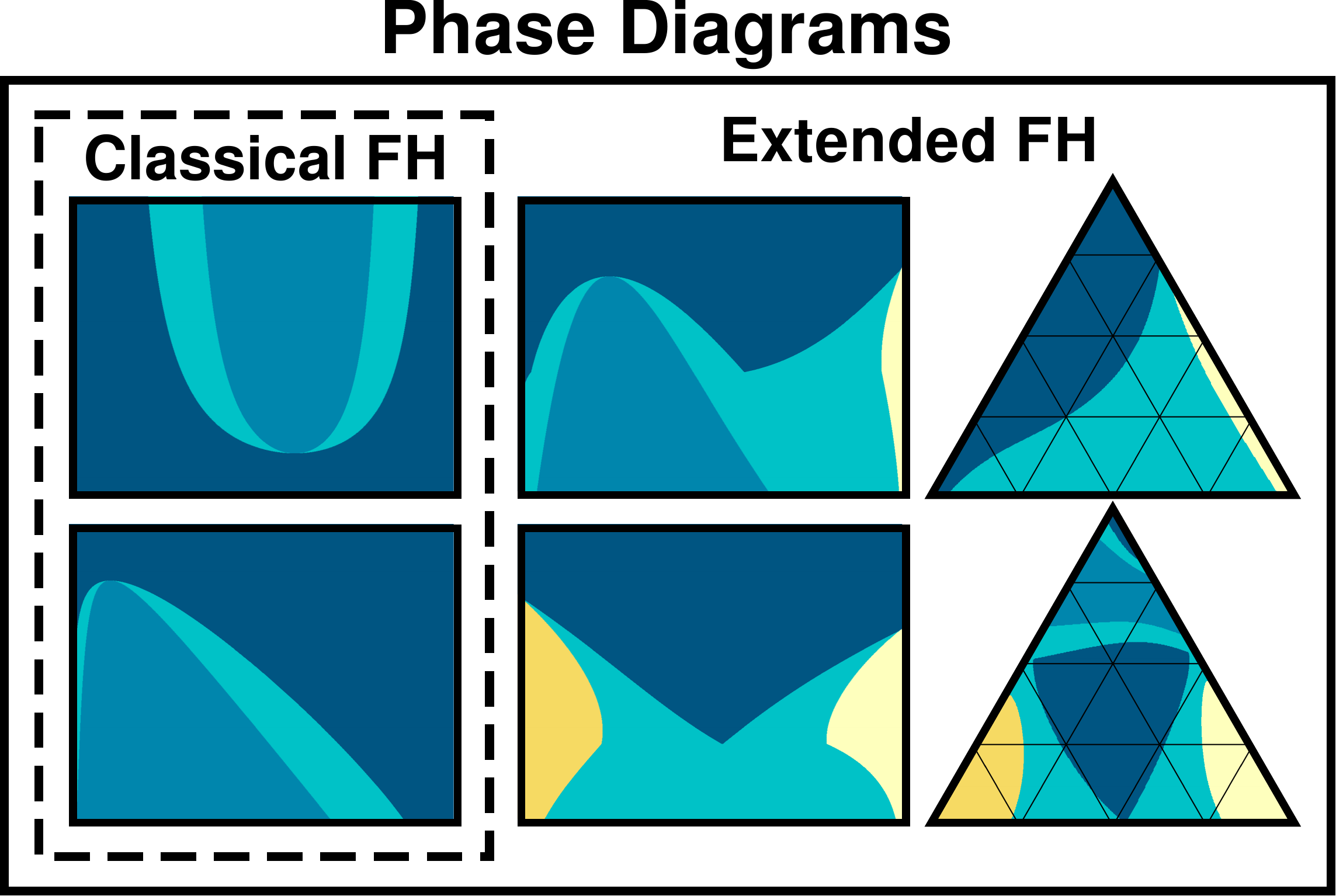}
    \caption{Graphical abstract}
    \label{fig:GraphicalAbstract}
\end{figure*}




\section{Abstract}\label{Sec:Abstract}

The Flory-Huggins theory is a well-established lattice model that is commonly used to study the mixing of distinct chemical species. It can successfully predict phase separation phenomena in blends of incompatible materials. However, it is limited to amorphous mixtures, excluding systems where the phase segregation is shaped by the concurrent crystallization of one or several blend components. A generalization of the Flory-Huggins formalism is thus necessary to capture the coupling and the interplay of crystallization with amorphous demixing mechanisms, such as spinodal decomposition. This work therefore revolves around the derivation of a free energy model for multicomponent mixtures that encompasses the physics of both processes. It is detailed which concepts from the original Flory-Huggins theory are required to apprehend the presented developments and how the current framework is built upon them. Furthermore, additional discussion points address chemical potential calculations and selected examples of binary and ternary phase diagrams, thereby highlighting the variety of blend behaviors that can be represented. 


\section{Introduction}\label{Sec:Intro}

The Flory-Huggins theory~\cite{flory_principles_1953,huggins_theory_1942} provides a mathematical formalism that describes the thermodynamics of material mixtures. It relies on a virtual lattice representation, which allows to evaluate the spatial arrangements of chemical species of different sizes, such as polymers and small molecules, for instance. The model was first developed for binary systems, but generalizations for any amorphous blend, regardless of the number of components, are usually employed as well~\cite{flory_principles_1953,hsu_thermodynamics_1974,boom_equilibrium_1994,horst_phase_1995-1,favre_application_1996-1,xu_simultaneous_2014,aryanti_flory-huggins_2018,mao_phase_2019,romay_thermodynamic_2021,ronsin_phase-field_2022}. 

Predictions from the Flory-Huggins framework were demonstrated to agree with qualitative observations from experiments~\cite{flory_principles_1953,rubinstein_polymer_2003}. Especially, mixtures that exhibit miscibility gaps and are prone to spinodal decomposition behavior are accounted for. Quantitatively, the theoretical expectations from the Flory-Huggins free energy model are in line with measurements~\cite{flory_principles_1953,favre_application_1996-1,jung_polymerization_2010,venetsanos_mixing_2022-1,ameslon_taxonomy_2025}, even though corrections have to be implemented for material combinations where mixing interactions display complex dependencies on temperature, composition, and chemical structure~\cite{petri_composition-dependent_1995,favre_application_1996-1,tambasco_blend_2006,nedoma_measurements_2008,loo_composition_2019,willis_simple_2020-1,romay_thermodynamic_2021}.

The theory finds applications in various research areas such as drug-polymer systems\cite{knopp_comparative_2015,donnelly_probing_2015,potter_investigation_2018,thakore_analytical_2021}, organic electronics~\cite{ghasemi_molecular_2021,ronsin_formation_2022,peng_materials_2023-1,kim_phase_2025}, and polymeric membrane manufacturing~\cite{boom_equilibrium_1994,barzin_theoretical_2007,aryanti_flory-huggins_2018, romay_thermodynamic_2021}, for example. Recent efforts have been dedicated to determine analytical solutions for the binodal equilibrium compositions predicted by the model, so as to facilitate its usage~\cite{qian_analytical_2022-1,de_souza_exact_2024}. A remaining limitation of the treatment of mixing in the Flory-Huggins free energy framework is its restriction to amorphous components, while many materials can undergo crystallization phase transitions, even in the blend.

The objective of this work is therefore to derive an extended model that is based on the classical Flory-Huggins theory and captures crystallization phenomena. For this purpose, the proposed approach follows a generalized version of the mean-field approximation that is usually applied to the enthalpic mixing interactions in the fully amorphous case~\cite{rubinstein_polymer_2003}. Some introduced features are also inspired by previous publications by Matkar and Kyu~\cite{matkar_phase_2006,matkar_role_2006} where the Flory-Huggins formalism was augmented with elements from the Landau theory for phase transitions~\cite{hohenberg_introduction_2015-1} and Phase-Field modeling~\cite{takaki_phase-field_2014,granasy_phase-field_2019,ronsin_phase-field_2022,siber_crystalline_2023} to obtain an expression for the free energy density of crystallization in binary mixtures. In addition, the common assumption that the latent heat release accompanying crystallization is linear with the degree of undercooling~\cite{turnbull_formation_1950} (as, for example, in the treatment of polymer crystallization by Hoffman and Lauritzen~\cite{lauritzen_theory_1960}) is used here as well. 

Following this introduction, the present manuscript is divided into four successive sections: The first one contains an overview of the aspects of the original theory that subsequent developments are built upon. The second then details the derivation of the free energy formula for multicomponent blends with any number of crystalline constituents. The third discusses further chemical potential calculations and features a showcase study of phase diagrams generated from the model. Finally, the fourth section exposes the conclusions of this work.


\section{Free Energy in the Classical Flory-Huggins Framework}\label{Sec:ClassicalFL}

\begin{figure}[H]
    \centering
    \includegraphics[scale=0.23]{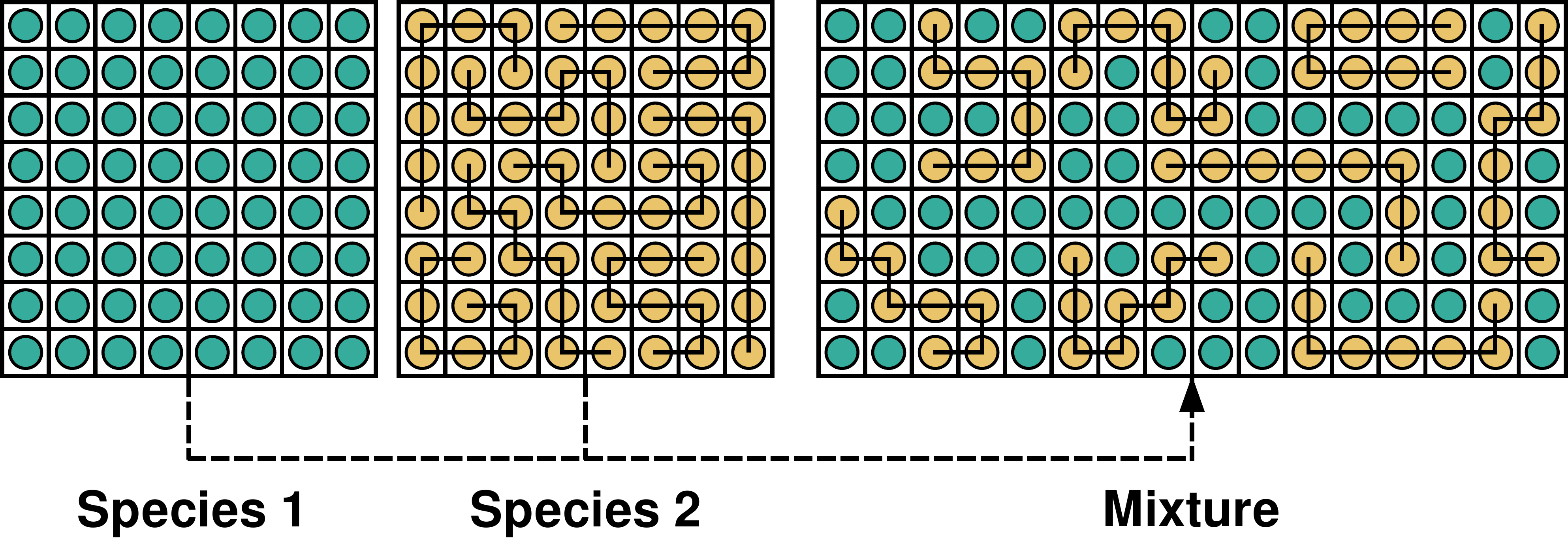}
    \caption{Schematic illustration of the mixing of a polymer solution on a two-dimensional Flory-Huggins lattice. The size proportions are $N_2=8$ for the polymer (in yellow) to $N_1 = 1$ for the solvent (in green).}
    \label{fig:FHMixing}
\end{figure}

This section reviews core concepts from the classical Flory-Huggins theory. The purpose is not to detail the derivation of the fundamental model equations, as this can readily be found elsewhere in the literature~\cite{flory_principles_1953,rubinstein_polymer_2003}, but rather to provide a reminder of the theoretical framework that the upcoming developments rely on. In this context, a material mixture is viewed as the arrangement of its different chemical constituents on a virtual lattice, hereafter referred to as the Flory-Huggins lattice (see Fig.~\ref{fig:FHMixing}). As a result, each component can occupy one or several sites of the lattice, depending on its size compared to the reference volume of a grid element. For convenience, the size of the smallest species in the blend is usually taken to determine the dimension of the Flory-Huggins lattice. In principle, the reference volume of a lattice element is, however, arbitrary. Despite having originally been built for binary polymer blends, the theory can be extended for amorphous mixtures with any number of components~\cite{flory_principles_1953,aryanti_flory-huggins_2018,mao_phase_2019,ronsin_phase-field_2022,ronsin_formation_2022}. For simplicity, the focus of this summary is restricted to mixtures involving two species only. 

The advantage of the Flory-Huggins approach lies in its relatively simple formulation for the system's free energy change upon mixing $\Delta G$, namely

\begin{equation}\label{eq:Sec1_FreeEnergyk}
    \Delta G = kT \left[\overline{n}_1 \ln{(\phi_1)} + \overline{n}_2 \ln{(\phi_2)} +  \overline{n}_0 \phi_1  \phi_2 \chi_{12} \vphantom{n_2^{(M)}} \right]~.
\end{equation}
Here, $k$ is the Boltzmann constant and $T$ the temperature. $\overline{n}_0$ represents the total number of Flory-Huggins lattice sites. Analogously, $\overline{n}_1$ and $\overline{n}_2$ denote the number of particles for the first and the second species, respectively. Their overall volume fractions are then symbolized by $\phi_1$ and $\phi_2$. The terms involving the logarithms describe the free energy change due to the entropy increase upon ideal mixing and are consistent with the predictions from regular solution theory~\cite{hildebrand_models_1953,hildebrand_history_1981,blanks_thermodynamics_1964}. 
Eq.~\ref{eq:Sec1_FreeEnergyk} slightly differs from the equation originally presented by Flory~\cite{flory_principles_1953} because it does not consider the Flory-Huggins lattice site volume to be necessarily equal to volume of the smallest species in the blend, as discussed in the supporting information (SI-A). Note also that $\Delta G$ conventionally refers to the Gibbs free energy. The formula for the Helmholtz counterpart ($\Delta F$) is nonetheless equivalent, the theory assuming that mixing occurs under constant volume and pressure conditions. 

In order to capture material mixtures that deviate from ideal mixing behavior, the Flory-Huggins model also includes a contribution that is controlled by the parameter $\chi_{12}$. This parameter accounts for either repulsive or attractive interactions arising between pairs of different species that occupy nearest-neighbor sites on the Flory-Huggins lattice~\cite{flory_principles_1953,rubinstein_polymer_2003}. It is defined as
\begin{equation}\label{eq:Sec1_InteractionParam}
    \chi_{12} := \frac{z}{kT} \Delta w = \frac{z}{kT} \left[w_{12} - \frac{1}{2} \left( w_{11} + w_{22} \right) \right]  ~,
\end{equation}
where $\Delta w$ is the energy gain per nearest-neighbor contact. $\Delta w$ originates from the interaction energy of a contact between both species $w_{12}$, which, upon mixing, replaces the interaction energy of a component with itself ($w_{11}$ or $w_{22}$)~\cite{flory_principles_1953,hillert_phase_2008}. In addition, $z$ is the so-called coordination number, that is the number of nearest neighbors to a site of the lattice.
In the most general case, $\Delta w$ is assumed to be of the same nature as a free energy, hence being decomposable in an enthalpy (or internal energy) and an entropy part. Empirically, the variation of $\chi_{12}$ with temperature is indeed found to obey the following relationship in many situations~\cite{rubinstein_polymer_2003}:
\begin{equation}\label{eq:Sec1_InteractionParamLinForm}
    \chi_{12} = A + \frac{B}{T} ~.
\end{equation}
$A$ and $B$ are the constant coefficients of the entropic and enthalpic contributions, respectively. Nevertheless, cases exist where this linear form in $1/T$ is not fitting the measurements \cite{loo_composition_2019}. Additionally, $\chi_{12}$ is overall expected to be composition-dependent~\cite{petri_composition-dependent_1995,rubinstein_polymer_2003,nedoma_measurements_2008,loo_composition_2019}, even though this is not directly addressed within the framework of the classical Flory-Huggins theory.

It has to be pointed out that the value of the interaction parameter changes with the reference size chosen for the grid elements of the lattice (see SI-A). A quantity that describes the miscibility of a specific material pair is rather the ratio between $\chi_{12}$ and the molar volume of the Flory-Huggins lattice sites $v_0$. To explain this further, one may consider the free energy density $\Delta G_V$ which, being an intensive quantity, does not change with the total volume of the system. To express $\Delta G_V$, $\Delta G$ can first be adapted in order to reason in terms of number of moles,
\begin{equation}\label{eq:Sec1_FreeEnergyR}
    \Delta G = RT \left[n_1\ln{(\phi_1)} + n_2\ln{(\phi_2)} + n_0 \phi_1 \phi_2 \vphantom{n_2^{(M)}} \chi_{12}  \right] ~,
\end{equation}
with $n_0$ the mole number of lattice sites and $n_1$ and $n_2$ the mole numbers of species $1$ and $2$, respectively. $\Delta G_V$ can then be obtained by dividing this latter equation by the volume of the mixture $V = n_0 v_0$ and substituting $n_1$ and $n_2$ by making use of relations between the mole numbers, the volume fractions, and the sizes of both components in terms of number of occupied lattice sites $N_1$ and $N_2$, that is 
\begin{equation}\label{eq:Sec1_SizeRelation}
    n_1 N_1 = n_0 \phi_1 \, , \, n_2 N_2 = n_0 \phi_2 ~.
\end{equation}
The free energy density $\Delta G_V$ reads 
\begin{equation}\label{eq:Sec1_FreeEnergyV}
   \Delta G_V = \frac{RT}{v_0} \left[ \frac{\phi_1}{N_1} \ln{(\phi_1)} + \frac{\phi_2}{N_2} \ln{(\phi_2)} + \phi_1 \phi_2 \chi_{12} \right] ~.
\end{equation}

Considering now the mixing of a same material blend projected onto two different Flory-Huggins lattices (denoted hereafter by the superscripts $(1)$ and $(2)$) with distinct lattice element sizes (i.e. reference molar volumes $v_0^{(1)}$ and $v_0^{(2)}$, respectively), the following equation immediately arises since the interaction energy contributions computed relatively to both reference systems ($RT\phi_1\phi_2\chi_{12}^{(1)}/v_0^{(1)}$ and $RT\phi_1\phi_2\chi_{12}^{(2)}/v_0^{(2)}$) are still required to be equal:

\begin{equation}\label{eq:Sec1_InteractionParamRefChange}
    \frac{\chi_{12}^{(1)}}{v_0^{(1)}} = \frac{\chi_{12}^{(2)}}{v_0^{(2)}} \Leftrightarrow \chi_{12}^{(2)} = \frac{v_0^{(2)}}{v_0^{(1)}} \chi_{12}^{(1)} ~.
\end{equation}

This provides a scaling relation that can be used to adapt the interaction parameter value from one reference lattice to another. Values of $\chi_{12}$ should therefore always be reported with the considered lattice molar volume $v_0$ in order to allow for reliable comparisons between different miscibility experiments.

Finally, formulae for the chemical potentials of both components can be obtained by taking the partial derivatives of the free energy (Eq.~\ref{eq:Sec1_FreeEnergyR}) with respect to the corresponding mole numbers $n_1$ and $n_2$. Note that the dependencies of the volume fractions $\phi_1$ and $\phi_2$ on $n_1$ and $n_2$ are taken into account during this calculation. Using Eq.~\ref{eq:Sec1_SizeRelation} and the fact that $\phi_1 + \phi_2 = 1$, the expressions of the chemical potentials $\mu_1$ and $\mu_2$ ultimately simplify to
\begin{equation}\label{eq:Sec1_ChemicalPotentials}
    \begin{dcases}
            \mu_1 = \frac{\partial \Delta G}{\partial n_1} = RT \left[ \ln{(\phi_1)}+ (1-\phi_1)\left(1- \frac{N_1}{N_2} \right) + N_1 \chi_{12} (1-\phi_1)^2 \right] ~, \\
            \mu_2 = \frac{\partial \Delta G}{\partial n_2} = RT \left[ \ln{(\phi_2)} + (1-\phi_2)\left(1 - \frac{N_2}{N_1} \right) + N_2 \chi_{12} (1-\phi_2)^2 \right] ~.
    \end{dcases}
\end{equation}

Again, as for Eq.~\ref{eq:Sec1_FreeEnergyk}, it can be observed that the original treatment by Flory~\cite{flory_principles_1953} results in slightly different chemical potentials due to the implicit scaling of the grid elements with the smallest component of the mixture (see SI-A for more details).

\section{Generalization for Crystalline Multicomponent Mixtures}\label{Sec:Derivation}

After having presented the features and equations of the classical Flory-Huggins theory that are fundamental for the present endeavour, the current section addresses the generalization of the model for material blends with any number of amorphous and/or crystalline components. Conceptually, the followed approach is analogous to the method detailed by Rubinstein and Colby.~\cite{rubinstein_polymer_2003} for the enthalpy part of the interaction parameter. In the present development, it is applied directly to the overall free energy of the system, which includes all possible enthalpic and entropic contributions. As compared to the original treatment, this implies the following supplementary assumptions:
\begin{enumerate}
    \item In addition to the enthalpy, the global entropy of the system (at equilibrium) can  be decomposed into a sum of respective entropic contributions from each pair of neighboring lattice sites that constitute the blend.
    \item The free energy gain associated with crystallization (i.e. the overall latent heat release and the surface tension arising at interfaces between crystalline and amorphous phases) can be modelled as resulting from bonding and repulsion interactions between nearest neighbors on the Flory-Huggins lattice, similarly as in the mean-field approach that yields the definition of the classical interaction parameter (Eq.~\ref{eq:Sec1_InteractionParam}). 
\end{enumerate}

Conventionally, the free energy change upon mixing $\Delta G$ is written as
\begin{equation}\label{eq:Sec2_GDiff}
    \Delta G = G - G^{(0)}~.
\end{equation}
Here, $G$ denotes the free energy after the mixing and crystallization processes have occurred and $G^{(0)}$ is the reference free energy of the system in the unmixed amorphous state. Being extensive quantities, these total free energies can be calculated by adding up all individual free energy contributions from the different grid elements that form the Flory-Huggins lattice. For convenience, this can be expressed in factorized forms as  
\begin{equation}\label{eq:Sec2_WeightedSumG}
\begin{dcases}
    G = \sum_{i=1}^n \phi_i \left[ (1-\psi_i) G_i^{(a)} + \psi_i G_i^{(c)} \right]~, \\
    G^{(0)} = \sum_{i=1}^n \phi_i  G_i^{(0)} ~.
\end{dcases}
\end{equation}
The summation is over the $n$ components of the blend and the free energy contributions pertaining to lattice sites filled with a given species $i$ are weighted by its corresponding total volume fraction $\phi_i$. In the formula for $G$, it is additionally distinguished whether the elements are in the crystalline or in the amorphous state (see superscript indices $(c)$ and $(a)$, respectively). A supplementary variable $\psi_i$ is introduced here to represent the relative crystallinity of species $i$. In this way, the products $\phi_i \psi_i$ and $\phi_i (1-\psi_i)$ give the volume fractions of crystalline and amorphous material $i$, respectively. These terms multiply the free energies $G_{i}^{(c)}$ and $G_{i}^{(a)}$ that belong to crystalline and amorphous lattice sites, so that their proportions in the blend are respected in the free energy formula. In comparison to $G$, a unique free energy contribution per species ($G_i^{(0)}$) is necessary to compute $G^{(0)}$ since all the components are still amorphous in the reference state. 

The lattice site contributions $G_i^{(a)}$, $G_i^{(c)}$, and $G_i^{(0)}$ can now be developed further following the assumption that they arise from nearest-neighbor interactions: 
\begin{equation}\label{eq:Sec2_PairwiseInt}
    \begin{dcases}
        G_i^{(a)} = \frac{n_0 N_A z}{2} \sum_{j=1}^n \phi_j \left[(1-\psi_j) G_{ij}^{(aa)} + \psi_j G_{ij}^{(ac)} \right] ~, \\
        G_i^{(c)} = \frac{n_0 N_A z}{2} \sum_{j=1}^n \phi_j \left[(1-\psi_j) G_{ij}^{(ca)} + \psi_j G_{ij}^{(cc)} \right] ~, \\
        G_i^{(0)} = \frac{n_0 N_A z}{2} \sum_{j=1}^n \delta_{ij} G_{ij}^{(aa)} = \frac{n_0 N_A z}{2} G_{ii}^{(aa)} ~.
    \end{dcases}
\end{equation}

For any site neighbor to the one filled with $i$ (for which either $G_i^{(a)}$, $G_i^{(c)}$, or $G_i^{(0)}$ is written), and occupied by species $j$ (which can be any of the $n$ constituents, $i$ included), four different types of interactions can occur depending on the state of both elements. The possible pair combinations are amorphous-amorphous, amorphous-crystalline, crystalline-amorphous, or crystalline-crystalline. For each one, a corresponding free energy contribution is considered: $G_{ij}^{(aa)}$, $G_{ij}^{(ac)}$, $G_{ij}^{(ca)}$, and $G_{ij}^{(cc)}$. The subscript indices ($i$, $j$) refer to the components involved in the interaction while the superscript indices ($a$, $c$) specify their respective state. The distinction between the amorphous-crystalline and crystalline-amorphous contributions ($G_{ij}^{(ac)}$ and $G_{ij}^{(ca)}$) matters, since these are a priori not necessarily symmetric. Employing the same mean-field treatment as in the classical Flory-Huggins theory~\cite{rubinstein_polymer_2003}, these terms are weighted by the probability of encountering the associated nearest-neighbor couple, knowing already that the lattice site occupied by component $i$ is involved in the pair.

Supposing isotropic mixing, the probability to have species $j$ neighboring species $i$ is its overall volume fraction $\phi_j$. In addition, the probability for $j$ to be amorphous is $1-\psi_j$ and $\psi_j$ to be crystalline. As in Eq.~\ref{eq:Sec2_WeightedSumG}, the free energy contributions $G_{ij}^{(aa)}$ and $G_{ij}^{(ca)}$ are thus scaled with $\phi_j (1-\psi_j)$, while $G_{ij}^{(ac)}$ and $G_{ij}^{(cc)}$ are multiplied by $\phi_j \psi_j$. This however neglects possible restrictions of the mixing due to preferential component arrangements induced by the crystallization, which may impact the expression of the aforementioned probabilities. Accounting for this is outside of the scope of this work, as more elaborated studies of the specific lattice conformations produced by investigated material combinations upon crystallization are necessary to refine the present model with this consideration. In the premixing configuration described by $G_i^{(0)}$, the probability to find component $j$ next to component $i$ can be expressed by the Kronecker symbol $\delta_{ij}$ since the constituents are only in contact with themselves. Moreover, the system is fully amorphous in this case, so only the $G_{ij}^{(aa)}$ contribution is active. Eq.~\ref{eq:Sec2_PairwiseInt} is finally obtained by summing over the number of components $n$, multiplying by the number of neighbors per lattice site $z$ times the total number of Flory-Huggins lattice elements $\overline{n}_0 = n_0 N_A$ (where $N_A$ denotes the Avogadro constant), and dividing by 2 to avoid counting twice each pairwise interaction.

Subsequently, Eq.~\ref{eq:Sec2_PairwiseInt} can be substituted into Eq.~\ref{eq:Sec2_WeightedSumG}:

\begin{equation}\label{eq:Sec2_WeightedSumGDeveloped}
    \begin{dcases}
        G = \frac{n_0 N_A z}{2} \sum_{i=1}^n \sum_{j=1}^n \phi_i \phi_j \left[(1-\psi_i)(1-\psi_j) G_{ij}^{(aa)} + (1-\psi_i)\psi_j G_{ij}^{(ac)} + \psi_i (1-\psi_j) G_{ij}^{(ca)} + \psi_i \psi_j G_{ij}^{(cc)} \right] ~, \\
        G^{(0)} = \frac{n_0 N_A z}{2} \sum_{i=1}^n \phi_i G_{ii}^{(aa)}~.
    \end{dcases}
\end{equation}

A major assumption of the Flory-Huggins theory is that the lattice is regular and invariant upon mixing~\cite{flory_principles_1953,rubinstein_polymer_2003}, so that the coordination number $z$ stays constant. Under this hypothesis, subtracting $G^{(0)}$ from $G$ leads to the relationship
\begin{equation}\label{eq:Sec2_FreeEnergyBasis}
\begin{split}
    \Delta G = \frac{n_0 N_A z}{2} & \left[ \vphantom{\sum_{j\neq i}^n} \sum_{i=1}^n \left( \phi_i^2 \left[(1-\psi_i)^2G_{ii}^{(aa)} + 2 \psi_i (1-\psi_i) G_{ii}^{(ac)} + \psi_i^2 G_{ii}^{(cc)} \right] - \phi_i G_{ii}^{(aa)} \right) + \right. \\ 
    & \left. \sum_{i=1}^n \sum_{j\neq i}^n \phi_i \phi_j \left[(1-\psi_i)(1-\psi_j) G_{ij}^{(aa)} + (1-\psi_i)\psi_j G_{ij}^{(ac)} + \psi_i (1-\psi_j) G_{ij}^{(ca)} + \psi_i \psi_j G_{ij}^{(cc)} \right] \right]~.
\end{split}
\end{equation}

In Eq.~\ref{eq:Sec2_FreeEnergyBasis}, contributions from neighbors of the same component are grouped separately to utilize the here existing symmetry between amorphous-crystalline and crystalline-amorphous interactions, that is $G_{ii}^{(ac)} = G_{ii}^{(ca)}$. From there, the aim is to recover the ideal mixing and interaction terms from the classical Flory-Huggins theory in order to ensure the consistency with the original model. First, the free energy contributions of the upper sum that involve crystalline elements are rewritten according to their deviation from the amorphous-amorphous interaction, i.e.~$G_{ii}^{(ac)} = G_{ii}^{(aa)} + \Delta G_{ii}^{(ac)}$ and $G_{ii}^{(cc)} = G_{ii}^{(aa)} + \Delta G_{ii}^{(cc)}$ (with $\Delta G_{ii}^{(ac)}$ and $\Delta G_{ii}^{(cc)}$ the respective correction terms for amorphous-crystalline and crystalline-crystalline contributions), so that
\begin{equation}\label{eq:Sec2_FreeEnergyRearrange1}
    \begin{split}
    \Delta G = \frac{n_0 N_A z}{2} & \left[ \vphantom{\sum_{j\neq i}^n} \sum_{i=1}^n \left( \phi_i^2 \left[ 2 \psi_i (1-\psi_i) \Delta G_{ii}^{(ac)} + \psi_i^2 \Delta G_{ii}^{(cc)} \right] - \phi_i (1-\phi_i) G_{ii}^{(aa)} \right) + \right. \\
    & \left. \sum_{i=1}^n \sum_{j\neq i}^n \phi_i \phi_j \left[(1-\psi_i)(1-\psi_j) G_{ij}^{(aa)} + (1-\psi_i)\psi_j G_{ij}^{(ac)} + \psi_i (1-\psi_j) G_{ij}^{(ca)} + \psi_i \psi_j G_{ij}^{(cc)} \right] \right]~.
\end{split}
\end{equation}

Utilizing the fact that adding up all volume fractions always amounts to 1, and thus that $1-\phi_i = \sum_{j\neq i}^n \phi_j$, it is possible to transfer the term in $G_{ii}^{(aa)}$ into the double sum. There, it may also be distributed as follows among the different contributions since $(1-\psi_i)(1-\psi_j) + (1-\psi_i)\psi_j + \psi_i(1-\psi_j) + \psi_i\psi_j = 1$:
\begin{equation}\label{eq:Sec2_FreeEnergyRearrange2}
\begin{split}
    \Delta G = \frac{n_0 N_A z}{2} & \left[ \vphantom{\sum_{j\neq i}^n} \sum_{i=1}^n  \phi_i^2 \left[ 2 \psi_i (1-\psi_i) \Delta G_{ii}^{(ac)} + \psi_i^2 \Delta G_{ii}^{(cc)} \right]  + \right. \\ 
    &  \sum_{i=1}^n \sum_{j\neq i}^n \phi_i \phi_j \left[  (1-\psi_i)(1-\psi_j) \left(G_{ij}^{(aa)}- G_{ii}^{(aa)}\right) + (1-\psi_i)\psi_j \left(G_{ij}^{(ac)}- G_{ii}^{(aa)}\right) \right. \\ 
    & \left. \vphantom{\sum_{j\neq i}^n} \left.  + \psi_i (1-\psi_j) \left(G_{ij}^{(ca)}- G_{ii}^{(aa)}\right) + \psi_i \psi_j \left(G_{ij}^{(cc)}- G_{ii}^{(aa)}\right) \right] \right]~.
\end{split}
\end{equation}

At this point, it is useful to express the free energy contributions of the second row in terms of their enthalpic (e.g. $H_{ij}^{(aa)}$) and entropic parts. The latter are additionally split into the entropy rise expected upon ideal mixing ($S_{ij}^{(id)}$) and a correction term (e.g. $\Delta S_{ij}^{(aa)}$) accounting for potential deviations from this behavior, yielding

\begin{equation}\label{eq:Sec2_EnergyContribDeveloped}
    \begin{dcases}
        G_{ij}^{(aa)} = H_{ij}^{(aa)} - T\left(S_{ij}^{(id)} + \Delta S_{ij}^{(aa)} \right) ~, \\
        G_{ij}^{(ac)} = H_{ij}^{(ac)} - T\left(S_{ij}^{(id)} + \Delta S_{ij}^{(ac)} \right) ~, \\
        G_{ij}^{(ca)} = H_{ij}^{(ca)} - T\left(S_{ij}^{(id)} + \Delta S_{ij}^{(ca)} \right) ~, \\
        G_{ij}^{(cc)} = H_{ij}^{(cc)} - T\left(S_{ij}^{(id)} + \Delta S_{ij}^{(cc)} \right) ~. \\
    \end{dcases}
\end{equation}

Replacing this into Eq.\ref{eq:Sec2_FreeEnergyRearrange2} and rearranging in order to regroup the ideal mixing entropies at the front of the double sum, one obtains
\begin{equation}\label{eq:Sec2_FreeEnergyRearrange3}
\begin{split}
    \Delta G = \frac{n_0 N_A z}{2} & \left[ \vphantom{\sum_{j\neq i}^n} \sum_{i=1}^n  \phi_i^2 \left[ 2 \psi_i (1-\psi_i) \Delta G_{ii}^{(ac)} + \psi_i^2 \Delta G_{ii}^{(cc)} \right]  + \sum_{i=1}^n \sum_{j\neq i}^n \phi_i \phi_j \left[ - T \left(S_{ij}^{(id)} - S_{ii}^{(id)} \right) \right. \right. \\ 
    & + \vphantom{\sum_{j\neq i}^n} (1-\psi_i)(1-\psi_j) \left(H_{ij}^{(aa)} - T \Delta S_{ij}^{(aa)} - H_{ii}^{(aa)} + T \Delta S_{ii}^{(aa)}\right) \\
    & + \vphantom{\sum_{j\neq i}^n} (1-\psi_i)\psi_j \left(H_{ij}^{(ac)} - T \Delta S_{ij}^{(ac)} - H_{ii}^{(aa)} + T \Delta S_{ii} ^{(aa)} \right) \\ 
    & + \vphantom{\sum_{j\neq i}^n} \psi_i (1-\psi_j) \left(H_{ij}^{(ca)} - T \Delta S_{ij}^{(ca)} - H_{ii}^{(aa)} + T \Delta S_{ii}^{(aa)} \right) \\ 
    & \left. \vphantom{\sum_{j\neq i}^n} \left. + \psi_i \psi_j \left(H_{ij}^{(cc)} - T \Delta S_{ij}^{(cc)} - H_{ii}^{(aa)} + T \Delta S_{ii}^{(aa)} \right) \right] \right]~.
\end{split}
\end{equation}

In the case of an ideal amorphous mixture, the free energy reduces to the latter mentioned entropy contributions, which must therefore identify with logarithmic terms comparable to those presented in Eq.~\ref{eq:Sec1_FreeEnergyR} to conform with the original model. Hence,
\begin{equation}\label{eq:Sec2_IdealEntropy}
   - \frac{T n_0 N_A z }{2} \sum_{i=1}^n \sum_{j\neq i}^n \phi_i \phi_j \left(S_{ij}^{(id)} - S_{ii}^{(id)} \right) = R T \sum_{i=1}^n n_i \ln{(\phi_i)}~.
\end{equation}

More details about the implied formulae for $S_{ij}^{(id)}$ and $S_{ii}^{(id)}$ can be found in the SI but are not necessary for the upcoming discussions (SI-B). Assuming all terms due to interactions between two different blend components $i$ and $j$ possess a symmetrical expression, i.e. $H_{ij}^{(aa)} - T\Delta S_{ij}^{(aa)} = H_{ji}^{(aa)} - T\Delta S_{ji}^{(aa)}$, $H_{ij}^{(ac)} - T\Delta S_{ij}^{(ac)}  = H_{ji}^{(ca)} - T \Delta S_{ji}^{(ca)}$, $H_{ij}^{(ca)} - T\Delta S_{ij}^{(ca)}  = H_{ji}^{(ac)} - T \Delta S_{ji}^{(ac)}$, and $H_{ij}^{(cc)} - T\Delta S_{ij}^{(cc)} = H_{ji}^{(cc)} - T\Delta S_{ji}^{(cc)}$, it can be observed that these actually appear twice in the remainder of the double sum. Thus, Eq.~\ref{eq:Sec2_FreeEnergyRearrange3} can equivalently be rewritten as 

\begin{equation}\label{eq:Sec2_FreeEnergyRearrange4}
\begin{split}
    \Delta G = & \frac{n_0 N_A z}{2}  \sum_{i=1}^n  \phi_i^2 \left[ 2 \psi_i (1-\psi_i) \Delta G_{ii}^{(ac)} + \psi_i^2 \Delta G_{ii}^{(cc)} \right]  + RT \sum_{i=1}^n n_i \ln{(\phi_i)} + \\ 
    & \frac{n_0 N_A z}{2} \sum_{i=1}^n\sum_{j > i}^n \left[ (1-\psi_i)(1-\psi_j) \left(2 H_{ij}^{(aa)} - 2 T \Delta S_{ij}^{(aa)} - H_{ii}^{(aa)} - H_{jj}^{(aa)} + T \Delta S_{ii}^{(aa)} + T \Delta S_{jj}^{(aa)}\right) \right. \\
    & + \vphantom{\sum_{j\neq i}^n} (1-\psi_i)\psi_j \left(2H_{ij}^{(ac)} - 2T \Delta S_{ij}^{(ac)} - H_{ii}^{(aa)} - H_{jj}^{(aa)} + T \Delta S_{ii}^{(aa)} + T \Delta S_{jj}^{(aa)}\right) \\ 
    & + \vphantom{\sum_{j\neq i}^n} \psi_i (1-\psi_j) \left(2H_{ij}^{(ca)} - 2T \Delta S_{ij}^{(ca)} - H_{ii}^{(aa)} - H_{jj}^{(aa)} + T \Delta S_{ii}^{(aa)} + T \Delta S_{jj}^{(aa)}\right) \\ 
    &  \vphantom{\sum_{j\neq i}^n} \left. + \psi_i \psi_j \left(2H_{ij}^{(cc)} - 2T \Delta S_{ij}^{(cc)} - H_{ii}^{(aa)} - H_{jj}^{(aa)} + T \Delta S_{ii}^{(aa)} + T \Delta S_{jj}^{(aa)}\right) \right]~.
\end{split}
\end{equation}

This allows to define interaction parameters for the four types of nearest-neighbor configurations that can occur (i.e., amorphous-amorphous, amorphous-crystalline, crystalline-amorphous, and crystalline-crystalline):
\begin{equation}\label{eq:Sec2_InteractionParams}
\begin{dcases}
     \chi_{ij}^{(aa)} := \frac{z}{kT} \left[H_{ij}^{(aa)} - \frac{1}{2}\left(H_{ii}^{(aa)} + H_{jj}^{(aa)} \right) - T \left( \Delta S_{ij}^{(aa)} - \frac{1}{2}  \left( \Delta S_{ii}^{(aa)} + \Delta S_{jj}^{(aa)} \right) \right)  \right] ~, \\
     \chi_{ij}^{(ac)} := \frac{z}{kT} \left[H_{ij}^{(ac)} - \frac{1}{2}\left(H_{ii}^{(aa)} + H_{jj}^{(aa)} \right) - T \left( \Delta S_{ij}^{(ac)} - \frac{1}{2}  \left( \Delta S_{ii}^{(aa)} + \Delta S_{jj}^{(aa)} \right) \right)  \right] ~, \\
     \chi_{ij}^{(ca)} := \frac{z}{kT} \left[H_{ij}^{(ca)} - \frac{1}{2}\left(H_{ii}^{(aa)} + H_{jj}^{(aa)} \right) - T \left( \Delta S_{ij}^{(ca)} - \frac{1}{2}  \left( \Delta S_{ii}^{(aa)} + \Delta S_{jj}^{(aa)} \right) \right)  \right] ~, \\
     \chi_{ij}^{(cc)} := \frac{z}{kT} \left[H_{ij}^{(cc)} - \frac{1}{2}\left(H_{ii}^{(aa)} + H_{jj}^{(aa)} \right) - T \left( \Delta S_{ij}^{(cc)} - \frac{1}{2}  \left( \Delta S_{ii}^{(aa)} + \Delta S_{jj}^{(aa)} \right) \right)  \right] ~.
\end{dcases}
\end{equation}
The first parameter, $\chi_{ij}^{(aa)}$, is analogous to the interaction parameter from the classical Flory-Huggins theory and directly takes the expected linear form in $1/T$ (see Eq.~\ref{eq:Sec1_InteractionParamLinForm}). As do the supplementary parameters $\chi_{ij}^{(ac)}$, $\chi_{ij}^{(ca)}$, and $\chi_{ij}^{(cc)}$, that stem from the extension for crystalline components. No explicit composition-dependencies arise from the present treatment, but it can be reminded that all the involved enthalpy and entropy terms may possibly be more complex functions of concentration, temperature, crystallinity, and material properties such as the polymer chain length, for example. Introducing the interaction parameters into Eq.~\ref{eq:Sec2_FreeEnergyRearrange4} results in
\begin{equation}\label{eq:Sec2_FreeEnergyRearrange5}
\begin{split}
    \Delta G = & \frac{n_0 N_A z}{2} \sum_{i=1}^n  \phi_i^2 \left[ 2 \psi_i (1-\psi_i) \Delta G_{ii}^{(ac)} + \psi_i^2 \Delta G_{ii}^{(cc)} \right]  + RT \sum_{i=1}^n n_i \ln{(\phi_i)} + \\ 
    &  n_0 RT \sum_{i=1}^n \sum_{j > i}^n  \phi_i \phi_j \left[ (1-\psi_i)(1-\psi_j)\chi_{ij}^{(aa)} + (1-\psi_i)\psi_j \chi_{ij}^{(ac)} + \psi_i (1-\psi_j) \chi_{ij}^{(ca)} + \psi_i \psi_j \chi_{ij}^{(cc)} \right] ~.
\end{split}
\end{equation}

Alternatively, it is also possible to describe the amorphous-crystalline, crystalline-amorphous, and crystalline-crystalline interactions relatively to the amorphous-amorphous ones via corresponding corrective parameters, namely  
\begin{equation}\label{eq:Sec2_CorrectionParams}
    \begin{dcases}
        \Delta \chi_{ij}^{(ac)} := \chi_{ij}^{(ac)} - \chi_{ij}^{(aa)} = \frac{z}{kT} \left[\Delta H_{ij}^{(ac)} - T \left( \Delta S_{ij}^{(ac)} - \Delta S_{ij}^{(aa)} \right)\right] ~, \\
        \Delta \chi_{ij}^{(ca)} := \chi_{ij}^{(ca)} - \chi_{ij}^{(aa)} = \frac{z}{kT} \left[\Delta H_{ij}^{(ca)} - T \left( \Delta S_{ij}^{(ca)} - \Delta S_{ij}^{(aa)} \right) \right] ~, \\
        \Delta \chi_{ij}^{(cc)} := \chi_{ij}^{(cc)} - \chi_{ij}^{(aa)} = \frac{z}{kT} \left[\Delta H_{ij}^{(cc)} - T \left( \Delta S_{ij}^{(cc)} - \Delta S_{ij}^{(aa)} \right) \right] ~,
    \end{dcases}
\end{equation}
with $\Delta H_{ij}^{(ac)} = H_{ij}^{(ac)} - H_{ij}^{(aa)}$, $\Delta H_{ij}^{(ca)} = H_{ij}^{(ca)} - H_{ij}^{(aa)}$, and $\Delta H_{ij}^{(cc)} = H_{ij}^{(cc)} - H_{ij}^{(aa)}$.

The total free energy upon mixing and crystallization then rather writes
\begin{equation}\label{eq:Sec2_FreeEnergyRearrange6}
\begin{split}
    \Delta G = & \frac{n_0 N_A z}{2} \sum_{i=1}^n  \phi_i^2 \left[ 2 \psi_i (1-\psi_i) \Delta G_{ii}^{(ac)} + \psi_i^2 \Delta G_{ii}^{(cc)} \right]  + RT \sum_{i=1}^n n_i \ln{(\phi_i)} \\ 
    &  + n_0 RT \sum_{i=1}^n \sum_{j > i}^n  \phi_i \phi_j \left[ \chi_{ij}^{(aa)} + (1-\psi_i)\psi_j \Delta \chi_{ij}^{(ac)} + \psi_i (1-\psi_j) \Delta \chi_{ij}^{(ca)} + \psi_i \psi_j \Delta \chi_{ij}^{(cc)} \right] ~.
\end{split}
\end{equation}

Note that, in either of both forms, all interaction parameters are still subjected to the scaling with the reference size of the Flory-Huggins lattice elements (see discussion around Eq.~\ref{eq:Sec1_InteractionParamRefChange} and SI-A).

All terms from the original theory are now recovered in Eq.~\ref{eq:Sec2_FreeEnergyRearrange5} and Eq.~\ref{eq:Sec2_FreeEnergyRearrange6}. The last steps of this derivation focus on incorporating as well the molar latent heats of the crystallizing species $\Delta h_i$, which are commonly employed in models of the crystallization phase transition~\cite{lauritzen_theory_1960,granasy_phase-field_2019}. In a fully crystallized one-component system, $\Delta G_{ii}^{(cc)}$ can for instance be related to $\Delta h_i$ by equating either Eq.~\ref{eq:Sec2_FreeEnergyRearrange5} or Eq.~\ref{eq:Sec2_FreeEnergyRearrange6} (reduced with $n=1$, $\psi_i = 1$, and $\phi_i = 1$, which implies here that $n_0 = n_i N_i $, $N_i$ standing for the size of species $i$ in terms of Flory-Huggins lattice elements) with the usual linear approximation of the crystallization free energy made in the vicinity of the equilibrium melting temperature $T_{m,i}$~\cite{turnbull_formation_1950,thompson_approximation_1979}: 

\begin{equation}\label{eq:Sec2_CrystHeat}
     \frac{n_0 N_A z}{2} \Delta G_{ii}^{(cc)} = n_i \Delta h_i \left(1-\frac{T}{T_{m,i}} \right) \Leftrightarrow \Delta G_{ii}^{(cc)} = \frac{2 \Delta h_i}{N_i N_A z} \left(1-\frac{T}{T_{m,i}} \right) ~.
\end{equation}

Moreover, one can also introduce a molar energy parameter $\Delta \sigma_i$ similar to $\Delta h_i$, so as to define $\Delta G_{ii}^{(ac)}$ as
\begin{equation}\label{eq:Sec2_CrystSurfTen}
    \Delta G_{ii}^{(ac)} := \frac{\Delta \sigma_i}{N_i N_A z} ~.
\end{equation}

This parameter controls the strength of the interaction between crystalline and amorphous regions of the same species, and is thus decisive for the surface tension between a crystal and its surrounding amorphous phase when the unary system is not completely crystallized. In accordance with expectations from crystallization modeling experiments~\cite{jian_temperature_2012,baidakov_temperature_2013,granasy_phase-field_2019}, it is assumed to bear a dependency on the system temperature. Even though its definition differs from that of the other interaction parameters, $\Delta \sigma_i$ has in fact a comparable nature. In effect, $\chi_{ij}^{(aa)}$ can also be shown to be responsible for surface tension properties that arise between two amorphous phases in a demixing-prone blend~\cite{vrij_equation_1968,enders_interfacial_1994}. In the same way, $\chi_{ij}^{(ac)}$, $\chi_{ij}^{(ca)}$, and $\chi_{ij}^{(cc)}$ (or $\Delta \chi_{ij}^{(ac)}$, $\Delta \chi_{ij}^{(ca)}$, and $\Delta \chi_{ij}^{(cc)}$) are anticipated be determinant for the surface energy at the corresponding amorphous-crystalline, crystalline-amorphous, and crystalline-crystalline interfaces. 

Under the hypothesis that $\Delta G_{ii}^{(cc)}$ and $\Delta G_{ii}^{(ac)}$ are constants at fixed temperature, Eq.~\ref{eq:Sec2_CrystHeat} and Eq.~\ref{eq:Sec2_CrystSurfTen} can readily be introduced in the free energy expressions (Eq.~\ref{eq:Sec2_FreeEnergyRearrange5} and Eq.~\ref{eq:Sec2_FreeEnergyRearrange6}), giving  
\begin{numcases}{}
\begin{split}
    \Delta G = & n_0 \sum_{i=1}^n  \frac{\phi_i^2}{N_i} \left[ \psi_i (1-\psi_i) \Delta \sigma_{i} + \psi_i^2 \Delta h_i\left(1-\frac{T}{T_{m,i}} \right) \right]  + RT \sum_{i=1}^n n_i \ln{(\phi_i)} + \\ 
    &  n_0 RT \sum_{i=1}^n \sum_{j > i}^n  \phi_i \phi_j \left[ (1-\psi_i)(1-\psi_j)\chi_{ij}^{(aa)} + (1-\psi_i)\psi_j \chi_{ij}^{(ac)} + \psi_i (1-\psi_j) \chi_{ij}^{(ca)} + \psi_i \psi_j \chi_{ij}^{(cc)} \right] ~,
\end{split}\label{eq:Sec2_FreeEnergyRearrange71} \\
\begin{split}
    \Delta G = & n_0 \sum_{i=1}^n  \frac{\phi_i^2}{N_i} \left[ \psi_i (1-\psi_i) \Delta \sigma_{i} + \psi_i^2 \Delta h_i\left(1-\frac{T}{T_{m,i}} \right) \right]  + RT \sum_{i=1}^n n_i \ln{(\phi_i)} + \\ 
    &  n_0 RT \sum_{i=1}^n \sum_{j > i}^n  \phi_i \phi_j \left[\chi_{ij}^{(aa)} + (1-\psi_i)\psi_j \Delta \chi_{ij}^{(ac)} + \psi_i (1-\psi_j) \Delta \chi_{ij}^{(ca)} + \psi_i \psi_j \Delta \chi_{ij}^{(cc)} \right] ~.
\end{split}\label{eq:Sec2_FreeEnergyRearrange72}
\end{numcases}

In the broadest case, it can nevertheless be expected that $\Delta G_{ii}^{(cc)}$ and $\Delta G_{ii}^{(ac)}$ exhibit more complex dependencies on blend composition, crystallinity, and material properties. More sophisticated crystallization models are then required to identify these terms with measurable quantities. Recalling that, in general, $n_0 \phi_i = n_i N_i$ (see Eq.~\ref{eq:Sec1_SizeRelation}), a final adjustment of the crystallization free energy is performed in Eq.~\ref{eq:Sec2_FreeEnergyRearrange71} and Eq.~\ref{eq:Sec2_FreeEnergyRearrange72}:
\begin{numcases}{}
\begin{split}
    \Delta G = & \sum_{i=1}^n  n_i \phi_i \left[ \psi_i (1-\psi_i) \Delta \sigma_{i} + \psi_i^2 \Delta h_i \left(1-\frac{T}{T_{m,i}} \right) \right]  + RT \sum_{i=1}^n n_i \ln{(\phi_i)} + \\ 
    &  n_0 RT \sum_{i=1}^n \sum_{j > i}^n  \phi_i \phi_j \left[ (1-\psi_i)(1-\psi_j)\chi_{ij}^{(aa)} + (1-\psi_i)\psi_j \chi_{ij}^{(ac)} + \psi_i (1-\psi_j) \chi_{ij}^{(ca)} + \psi_i \psi_j \chi_{ij}^{(cc)} \right] ~,
\end{split}\label{eq:Sec2_FreeEnergyRearrange81} \\
\begin{split}
    \Delta G = & \sum_{i=1}^n  n_i \phi_i \left[ \psi_i (1-\psi_i) \Delta \sigma_{i} + \psi_i^2 \Delta h_i \left(1-\frac{T}{T_{m,i}} \right) \right]  + RT \sum_{i=1}^n n_i \ln{(\phi_i)} + \\ 
    &  n_0 RT \sum_{i=1}^n \sum_{j > i}^n  \phi_i \phi_j \left[ \chi_{ij}^{(aa)} + (1-\psi_i)\psi_j \Delta \chi_{ij}^{(ac)} + \psi_i (1-\psi_j) \Delta \chi_{ij}^{(ca)} + \psi_i \psi_j \Delta \chi_{ij}^{(cc)} \right] ~.
\end{split}\label{eq:Sec2_FreeEnergyRearrange82}
\end{numcases}

\section{Chemical Potential and Phase Diagram Calculations}\label{Sec:Discussion}

The extended free energy also allows for chemical potentials $\mu_i = \partial \Delta G / \partial n_i$ to be calculated for each of the $n$ blend components, namely
%

\begin{equation}\label{eq:Sec3_ChemicalPotential1}
\begin{split}
    \mu_i = & \phi_i\left(2-\phi_i\right) \left[ \psi_i (1-\psi_i) \Delta \sigma_{i} + \psi_i^2 \Delta h_i \left(1-\frac{T}{T_{m,i}} \right) \right]  \\
    &  - \sum_{j \neq i}^n \left(\phi_j^2 \frac{N_i}{N_j} \left[ \psi_j (1-\psi_j) \Delta \sigma_{j} + \psi_j^2 \Delta h_j \left(1-\frac{T}{T_{m,j}} \right) \right] \right) + RT \left[ \vphantom{\sum_{\substack{k>j\\k \neq i}}^n} \ln{(\phi_i)} + \sum_{j \neq i}^n \phi_j \left(1 - \frac{N_i}{N_j} \right) \right. \\
    &  + N_i \sum_{j \neq i}^n \phi_j \left( \vphantom{\sum_{\substack{k>j\\k \neq i}}^n} \sum_{k \neq i}^n \phi_k \left[ (1-\psi_i)(1-\psi_k) \chi_{ik}^{(aa)} + (1-\psi_i)\psi_k \chi_{ik}^{(ac)} + \psi_i (1-\psi_k) \chi_{ik}^{(ca)} + \psi_i \psi_k \chi_{ik}^{(cc)} \right] \right. \\
    & \left. \left. - \sum_{\substack{k>j\\k \neq i}}^n \phi_k \left[ (1-\psi_j)(1-\psi_k) \chi_{jk}^{(aa)} + (1-\psi_j)\psi_k \chi_{jk}^{(ac)} + \psi_j (1-\psi_k) \chi_{jk}^{(ca)} + \psi_j \psi_k \chi_{jk}^{(cc)} \right] \right) \right] ~,
\end{split}
\end{equation}
or
\begin{equation}\label{eq:Sec3_ChemicalPotential2}
\begin{split}
    \mu_i = & \phi_i\left(2-\phi_i\right) \left[ \psi_i (1-\psi_i) \Delta \sigma_{i} + \psi_i^2 \Delta h_i \left(1-\frac{T}{T_{m,i}} \right) \right] \\
    & - \sum_{j \neq i}^n \left( \phi_j^2 \frac{N_i}{N_j} \left[ \psi_j (1-\psi_j) \Delta \sigma_{j} + \psi_j^2 \Delta h_j \left(1-\frac{T}{T_{m,j}} \right) \right] \right) + RT \left[ \vphantom{\sum_{\substack{k>j\\k \neq i}}^n} \ln{(\phi_i)} + \sum_{j \neq i}^n \phi_j \left(1 - \frac{N_i}{N_j} \right) \right. \\
    &  + N_i \sum_{j \neq i}^n \phi_j \left( \vphantom{\sum_{\substack{k>j\\k \neq i}}^n} \sum_{k \neq i}^n \phi_k \left[ \chi_{ik}^{(aa)} + (1-\psi_i)\psi_k \Delta \chi_{ik}^{(ac)} + \psi_i (1-\psi_k) \Delta \chi_{ik}^{(ca)} + \psi_i \psi_k \Delta \chi_{ik}^{(cc)} \right] \right. \\
    & \left. \left. - \sum_{\substack{k>j\\k \neq i}}^n \phi_k \left[ \chi_{jk}^{(aa)} + (1-\psi_j)\psi_k \Delta \chi_{jk}^{(ac)} + \psi_j (1-\psi_k) \Delta \chi_{jk}^{(ca)} + \psi_j \psi_k \Delta \chi_{jk}^{(cc)} \right] \right) \right] ~,
\end{split}
\end{equation}
depending on whether the first (Eq.~\ref{eq:Sec2_FreeEnergyRearrange81}) or the second (Eq.~\ref{eq:Sec2_FreeEnergyRearrange82}) form is considered for $\Delta G$ (see SI-C for the detailed chemical potential derivation). In the case of an amorphous binary system (where $n = 2$, $\psi_1 = \psi_2 = 0$, and $\phi_2 = 1-\phi_1$), it can be verified that these expressions reduce to Eq.~\ref{eq:Sec1_ChemicalPotentials}, as expected. 

Furthermore, for a two-component mixture in which solely one of the species is subject to crystallization and eventually forms a pure crystalline phase at equilibrium with a remaining mixed amorphous phase, it is possible to recover the melting point depression formula~\cite{flory_principles_1953} that can be utilized to assess the value of the classical Flory-Huggins interaction parameter from experimental liquidus measurements. For this, the chemical potential of the crystallizing species is evaluated with Eq.~\ref{eq:Sec3_ChemicalPotential1} (or Eq.~\ref{eq:Sec3_ChemicalPotential2}) separately in both phases. In what follows, it is assumed without loss of generality that species 1 is the crystallizing one. Moreover, quantities pertaining to the crystalline and the amorphous phase are referenced by the superscripts $(c)$ and $(a)$, respectively. Equating both chemical potentials (thus denoted by $\mu_1^{(c)}$ and $\mu_1^{(a)}$), one obtains
\begin{equation}\label{eq:Sec3_MeltingPointDepression}
    \mu_1^{(c)} = \mu_1^{(a)}  \Leftrightarrow \frac{\Delta h_1}{R} \left(\frac{1}{T}-\frac{1}{T_{m,1}} \right)=  \ln{(\phi_1^{(a)})}+ (1-\phi_1^{(a)})\left(1- \frac{N_1}{N_2} \right) + N_1 \chi_{12} (1-\phi_1^{(a)})^2 ~. 
\end{equation}

Note that the assumption that the crystalline phase is pure and perfectly ordered is crucial here because it implies that $\phi_1^{(c)} = 1$ and $\psi_1^{(c)} = 1$, so that ultimately $\mu_1^{(c)} = \Delta h_1 \left(1-T/T_{m,1} \right)$. Otherwise, the equation of the chemical potentials $\mu_1^{(c)}$ and $\mu_1^{(a)}$ becomes more complex and effects related to the crystalline-amorphous interaction parameter also come into play. Upon additional simplifications that differentiate between polymers and small molecules, this general formula (Eq.~\ref{eq:Sec3_MeltingPointDepression}) can further be transformed into the commonly employed relationships presented by Nishi and Wang~\cite{nishi_melting_1975}.

Besides predicting the melting point depression, the present free energy model can also be used to produce phase diagrams. To do so, it is convenient to rely on the free energy density:
\begin{numcases}{}
\begin{split}\label{eq:Sec3_FreeEnergyDensity1}
    \Delta G_V = & \sum_{i=1}^n  \frac{\phi_i^2}{v_i} \left[ \psi_i (1-\psi_i) \Delta \sigma_{i} + \psi_i^2 \Delta h_i \left(1-\frac{T}{T_{m,i}} \right) \right]  + \frac{RT}{v_0} \sum_{i=1}^n \frac{\phi_i}{N_i} \ln{(\phi_i)} \\ 
    &  + \frac{RT}{v_0} \sum_{i=1}^n \sum_{j > i}^n  \phi_i \phi_j \left[ (1-\psi_i)(1-\psi_j)\chi_{ij}^{(aa)} + (1-\psi_i)\psi_j \chi_{ij}^{(ac)} + \psi_i (1-\psi_j) \chi_{ij}^{(ca)} + \psi_i \psi_j \chi_{ij}^{(cc)} \right] ~,
\end{split} \\
\begin{split}\label{eq:Sec3_FreeEnergyDensity2}
    \Delta G_V = & \sum_{i=1}^n  \frac{\phi_i^2}{v_i} \left[ \psi_i (1-\psi_i) \Delta \sigma_{i} + \psi_i^2 \Delta h_i \left(1-\frac{T}{T_{m,i}} \right) \right]  + \frac{RT}{v_0} \sum_{i=1}^n \frac{\phi_i}{N_i} \ln{(\phi_i)} \\ 
    &  + \frac{RT}{v_0} \sum_{i=1}^n \sum_{j > i}^n  \phi_i \phi_j \left[ \chi_{ij}^{(aa)} + (1-\psi_i)\psi_j \Delta \chi_{ij}^{(ac)} + \psi_i (1-\psi_j) \Delta \chi_{ij}^{(ca)} + \psi_i \psi_j \Delta \chi_{ij}^{(cc)} \right] ~,
\end{split}
\end{numcases}
where $v_i = v_0 N_i$ stands for the molar volume of species $i$.

In this work, the convex hull approach~\cite{hildebrandt_predicting_1994-1,voskov_ternapi_2015,mao_phase_2019,gottl_convex_2023} is used to determine the different regions of the phase diagrams, although other methods exist as well~\cite{horst_calculation_1995,horst_calculation_1996}. All the diagrams presented hereafter are calculated from the second free energy form (Eq.~\ref{eq:Sec3_FreeEnergyDensity2}). Nonetheless, exactly the same figures can be achieved with the alternative form (Eq.~\ref{eq:Sec3_FreeEnergyDensity1}) and the adequate interaction parameters $\chi_{ij}^{(ac)}$, $\chi_{ij}^{(ca)}$, and $\chi_{ij}^{(cc)}$, instead of $\Delta \chi_{ij}^{(ac)}$, $\Delta \chi_{ij}^{(ca)}$, and $\Delta \chi_{ij}^{(cc)}$, respectively. The reason to rather use the second form over the first is that additional interactions involving crystalline components are considered relatively to the strength of the amorphous-amorphous ones, which facilitates the exploration of the parameter space. With the first formula, parameter combinations that cause atypical diagram shapes, and are not expected for most physical systems, are more likely to be encountered. For example, when employing moderate values of $\chi_{ij}^{(aa)}$ associated with relatively low $\chi_{ij}^{(ca)}$, $\chi_{ij}^{(ac)}$, and $\chi_{ij}^{(cc)}$, the free energy may favor a crystalline state at equilibrium, even without any crystallization driving force (i.e. at vanishing $\Delta h_i (1-T/T_{m,i})$). To obtain this with Eq.~\ref{eq:Sec3_FreeEnergyDensity2}, one would need to explicitly counter the magnitude of $\chi_{ij}^{(aa)}$ with accordingly negative correction parameters $\Delta \chi_{ij}^{(ac)}$, $\Delta \chi_{ij}^{(ca)}$, and $\Delta \chi_{ij}^{(cc)}$.

\begin{figure}[H]
    \centering
    \includegraphics[scale=0.5]{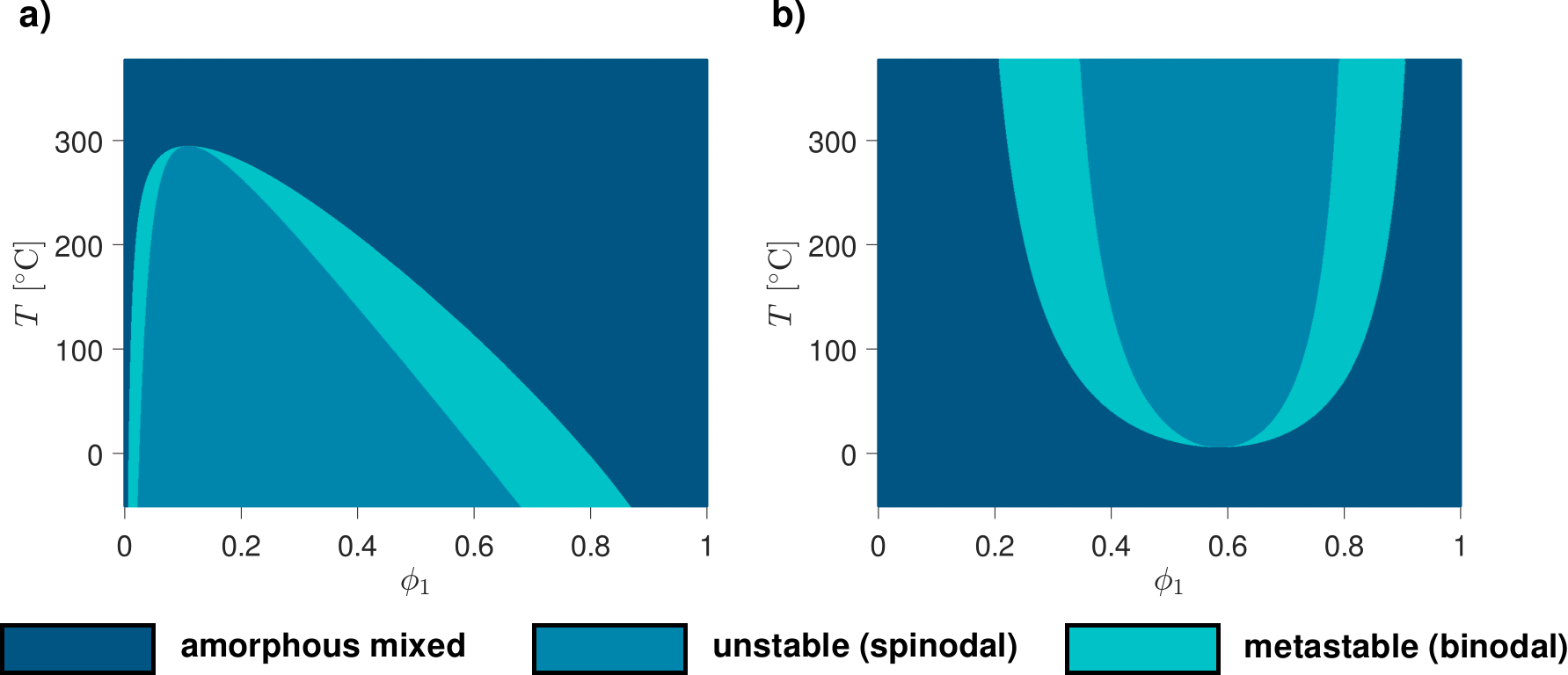}
    \caption{Phase diagrams of binary mixtures subject to a) UCST-type and b) LCST-type amorphous demixing. In a), the blend is strongly asymmetric ($N_1 = 100$ and $N_2 = 1$), which results in a miscibility gap that leans toward compositions richer in the smaller constituent. In b), the asymmetry is not as severe ($N_1 = 1$ and $N_2 = 2$) and the immiscible region is accordingly more centered. All relevant parameters used for the calculation of the diagrams are provided in the SI (SI-D).}
    \label{fig:BinaryAmorphous}
\end{figure}

\begin{figure}[H]
    \centering
    \includegraphics[scale=0.5]{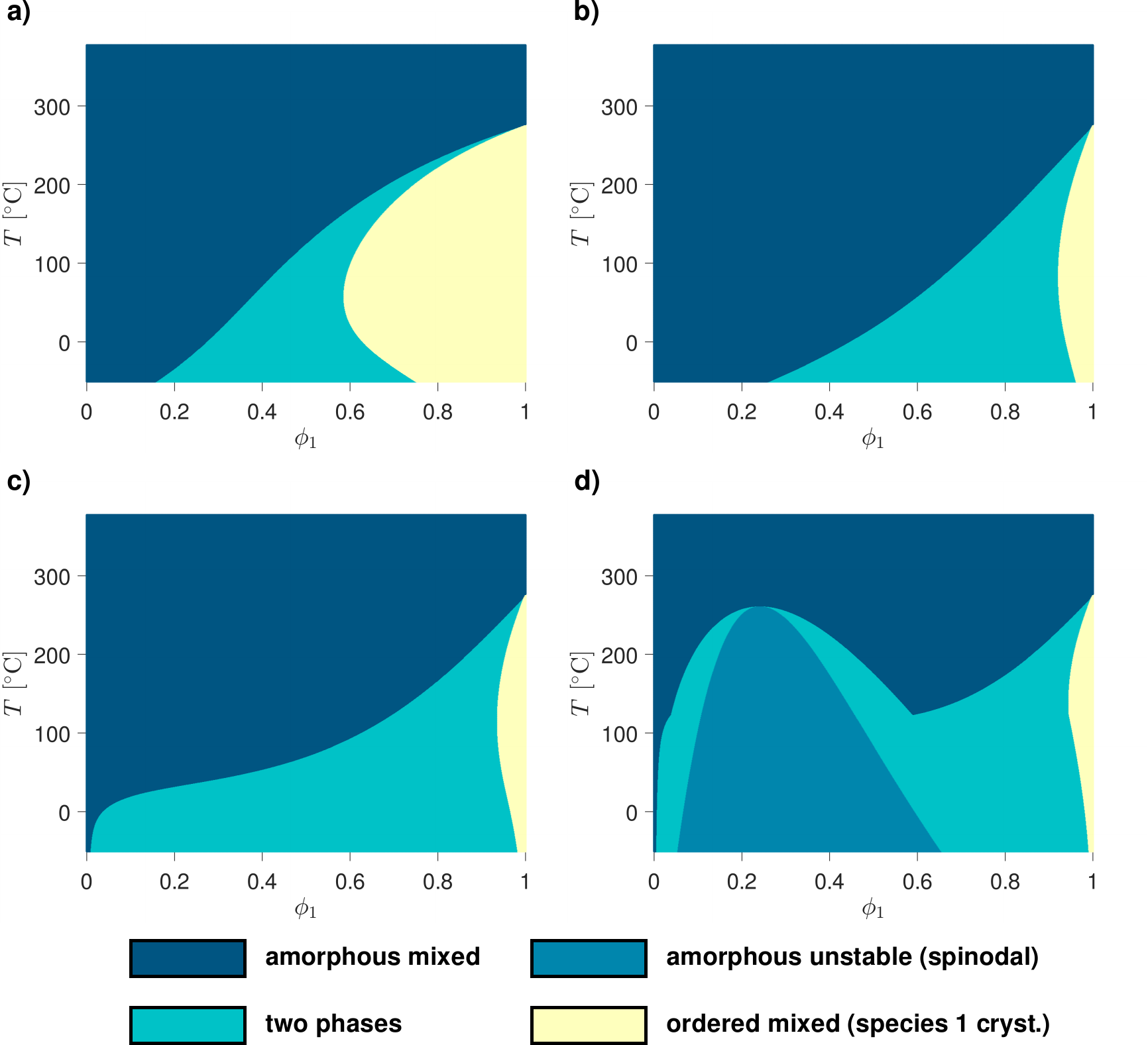}
    \caption{Phase diagrams of binary mixtures containing one species that can crystallize. In the first row, only the crystalline-amorphous parameter $\Delta \chi_{12}^{(ca)}$ is active and increased from a) $100/T$ to b) $0.2 + 350/T$. In the second row, $\Delta \chi_{12}^{(ca)}$ is maintained at $0.2 + 350/T$ while the effect of amorphous-amorphous interactions ranging from c) $\chi_{12}^{(aa)} = 0.3 + 110/T$ to d) $\chi_{12}^{(aa)} = 0.4 + 250/T$ is added. All relevant parameters used for the calculation of the diagrams are provided in the SI (SI-D).}
    \label{fig:Binary1Cryst}
\end{figure}

\begin{figure}[H]
    \centering
    \includegraphics[scale=0.5]{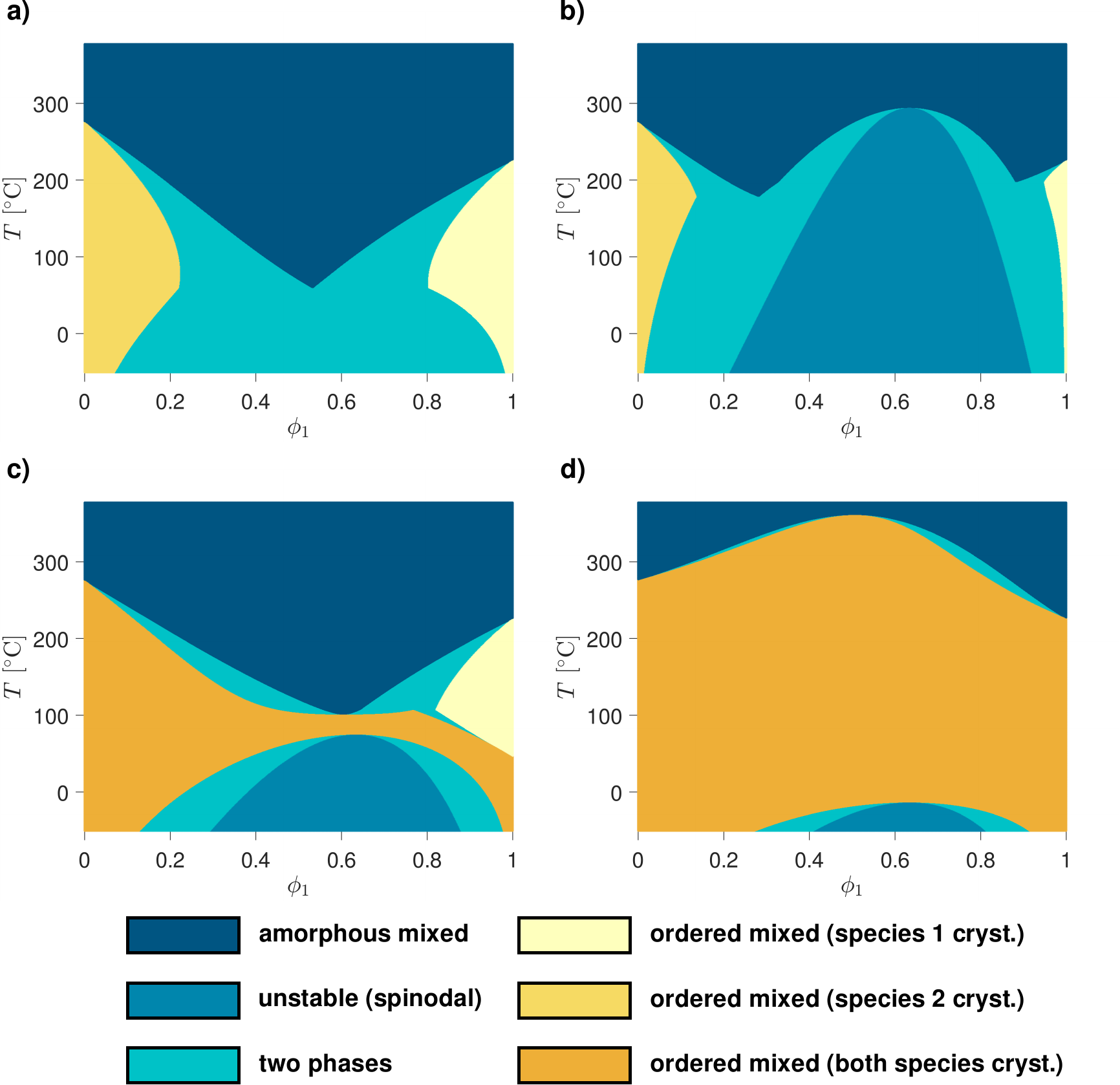}
    \caption{Phase diagrams of binary mixtures where both components can crystallize. In the first row, the amorphous-amorphous, amorphous-crystalline and crystalline-amorphous interaction parameters are active (i.e. $\chi_{12}^{(aa)}$, $\Delta \chi_{12}^{(ac)}$, and $\Delta \chi_{12}^{(ca)}$). $\chi_{12}^{(aa)}$ is increased from a) $0.3 + 150/T$ to b) $0.4 + 480/T$, while $\Delta \chi_{12}^{(ac)}$ and $\Delta \chi_{12}^{(ca)}$ are maintained constant at $0.1 + 100/T$ and $0.05 + 50/T$, respectively. In the second row, the effect of crystalline-crystalline compatibility is added. $\Delta \chi_{12}^{(cc)}$ is accordingly decreased from c) $0.3-30/T$ to d) $-60/T$. In both c) and d), the values of $\chi_{12}^{(aa)}$, $\Delta \chi_{12}^{(ac)}$, and $\Delta \chi_{12}^{(ca)}$ are the same as in a). All relevant parameters used for the calculation of the diagrams are provided in the SI (SI-D).}
    \label{fig:Binary2Cryst}
\end{figure}

Fig.~\ref{fig:BinaryAmorphous}, Fig.~\ref{fig:Binary1Cryst}, and Fig.~\ref{fig:Binary2Cryst} display typical examples of diagram shapes for binary systems. In Fig.~\ref{fig:BinaryAmorphous}, mixtures prone to amorphous demixing without any crystallization phase transition are modelled. The location of the spinodal and binodal gaps depends exclusively on the values of $\chi_{12}^{(aa)}$, $N_1$ and $N_2$, as already established within the framework of the classical Flory-Huggins theory~\cite{rubinstein_polymer_2003,konig_two-dimensional_2021}. Upper and lower critical solution temperature behavior (UCST and LCST) is obtained when using either a positive or a negative $B$ coefficient in the formula for $\chi_{12}^{(aa)}$ (Eq.~\ref{eq:Sec1_InteractionParamLinForm}). In addition, in the LCST case, $A$ has to be higher than the critical $\chi_{12}^{(aa)}$ value~\cite{rubinstein_polymer_2003,siber_crystalline_2023} above which the blend is susceptible to demix.

Fig.~\ref{fig:Binary1Cryst} shows diagrams for blends containing one crystallizing species. The effect of $\Delta \chi_{12}^{(ca)}$ is isolated in Fig.~\ref{fig:Binary1Cryst}-a and Fig.~\ref{fig:Binary1Cryst}-b. It can be seen that increasing the strength of the crystalline-amorphous interactions shifts the phase separation domain towards higher concentrations of the crystalline component. Additionally, the gap widens and the liquidus becomes concave at most volume fractions (except close to $\phi = 0$ and possibly $\phi = 1$). For most crystallizing mixtures, it is anticipated that both $\chi_{12}^{(aa)}$ and $\Delta \chi_{12}^{(ca)}$ are active. Adding an increasing $\chi_{12}^{(aa)}$ with a positive $B$ coefficient expands the two-phase region (Fig.~\ref{fig:Binary1Cryst}-c) up until an amorphous immiscibility region emerges above the liquidus (Fig.~\ref{fig:Binary1Cryst}-d).

Fig.~\ref{fig:Binary2Cryst} now addresses the situation where both components can crystallize. Fig.~\ref{fig:Binary2Cryst}-a demonstrates a typical diagram shape with a eutectic point obtained for moderate $\chi_{12}^{(aa)}$, $\Delta \chi_{12}^{(ac)}$, and $\Delta \chi_{12}^{(ca)}$ values. Fig.~\ref{fig:Binary2Cryst}-b then illustrates the possible interplay with an amorphous demixing region induced by a relatively high $\chi_{12}^{(aa)}$. In both Fig.~\ref{fig:Binary2Cryst}-a and Fig.~\ref{fig:Binary2Cryst}-b, no crystalline-crystalline interactions are considered, and the ordered phases arise from the crystallization of only one of the components. The other species may be mixed into that phase (for instance as defects or on interstitial sites of the crystal lattice) but does not explicitly form bonds and generate a latent heat release. 

Starting from the parameter set of Fig.~\ref{fig:Binary2Cryst}-a, Fig.~\ref{fig:Binary2Cryst}-c and Fig.~\ref{fig:Binary2Cryst}-d exemplify how the phase equilibria evolve when $\Delta \chi_{12}^{(cc)}$ becomes progressively more negative, that is the blended materials become increasingly more compatible in the crystalline state. It can be seen that this triggers the appearance of a region where both components contribute together to the crystallization process, thus forming co-crystals. Moreover, the slopes of the melting point depressions are damped (Fig.~\ref{fig:Binary2Cryst}-c) and ultimately inverted, leading to higher melting temperatures in the blend as compared to the pure materials (Fig.~\ref{fig:Binary2Cryst}-d). It can also be remarked that these diagrams predict a miscibility gap where spinodal/binodal phase separation takes place in the ordered state.

Comparing the diagram types produced from this model with the prior one of Matkar and Kyu~\cite{matkar_phase_2006}, it can be seen that both lead to similar features. A notable distinction is that diagrams computed from the model of Matkar and Kyu tend to exhibit fully ordered phases below a certain threshold temperature, even at vanishing content of the actual crystallizing components, as depicted in the SI (SI-E). This feature is not anticipated for most physical systems and is also not witnessed with the current free energy.


Another qualitative difference concerns the mathematical form of the crystallization energy (see SI-E). With the present formulation, it varies with the square of the crystallinity. In contrast, the framework of Matkar and Kyu relies on a Landau expansion~\cite{hohenberg_introduction_2015-1} which employs polynomials of higher order. It is shown in the SI how this latter approach can also be incorporated into the formulae developed here (SI-E), resulting in the following free energy densities:
\begin{numcases}{}
\begin{split}\label{eq:Sec3_FreeEnergyDensity3}
    \Delta G_V = & \sum_{i=1}^n  \frac{\phi_i^2}{v_i} \left[ \psi_i^2 (1-\psi_i)^2 \Delta \sigma_{i} + \psi_i^2(3-2\psi_i) \Delta h_i \left(1-\frac{T}{T_{m,i}} \right) \right]  + \frac{RT}{v_0} \sum_{i=1}^n \frac{\phi_i}{N_i} \ln{(\phi_i)} \\ 
    &  + \frac{RT}{v_0} \sum_{i=1}^n \sum_{j > i}^n  \phi_i \phi_j \left[ (1-\psi_i)(1-\psi_j)\chi_{ij}^{(aa)} + (1-\psi_i)\psi_j \chi_{ij}^{(ac)} + \psi_i (1-\psi_j) \chi_{ij}^{(ca)} + \psi_i \psi_j \chi_{ij}^{(cc)} \right] ~,
\end{split} \\
\begin{split}\label{eq:Sec3_FreeEnergyDensity4}
    \Delta G_V = & \sum_{i=1}^n  \frac{\phi_i^2}{v_i} \left[ \psi_i^2 (1-\psi_i)^2 \Delta \sigma_{i} + \psi_i^2 (3-2\psi_i) \Delta h_i \left(1-\frac{T}{T_{m,i}} \right) \right]  + \frac{RT}{v_0} \sum_{i=1}^n \frac{\phi_i}{N_i} \ln{(\phi_i)} \\ 
    &  + \frac{RT}{v_0} \sum_{i=1}^n \sum_{j > i}^n  \phi_i \phi_j \left[ \chi_{ij}^{(aa)} + (1-\psi_i)\psi_j \Delta \chi_{ij}^{(ac)} + \psi_i (1-\psi_j) \Delta \chi_{ij}^{(ca)} + \psi_i \psi_j \Delta \chi_{ij}^{(cc)} \right] ~.
\end{split}
\end{numcases}

Phase diagrams generated from these expressions possess qualitative properties comparable to those already presented and are therefore not discussed. Nevertheless, the free energy forms stemming from the Landau theory predict a so-called ``spinodal temperature''~\cite{granasy_phase-field_2019} below which the crystallization energy barrier vanishes, so that the phase transition may proceed spontaneously without following a nucleation and growth process. This is not the case with the current free energy density (Eq.~\ref{eq:Sec3_FreeEnergyDensity1} and Eq.~\ref{eq:Sec3_FreeEnergyDensity2}), where the barrier always exists until $T = 0$~K (see discussion in SI-E). 



\begin{figure}[H]
    \centering
    \includegraphics[scale=0.48]{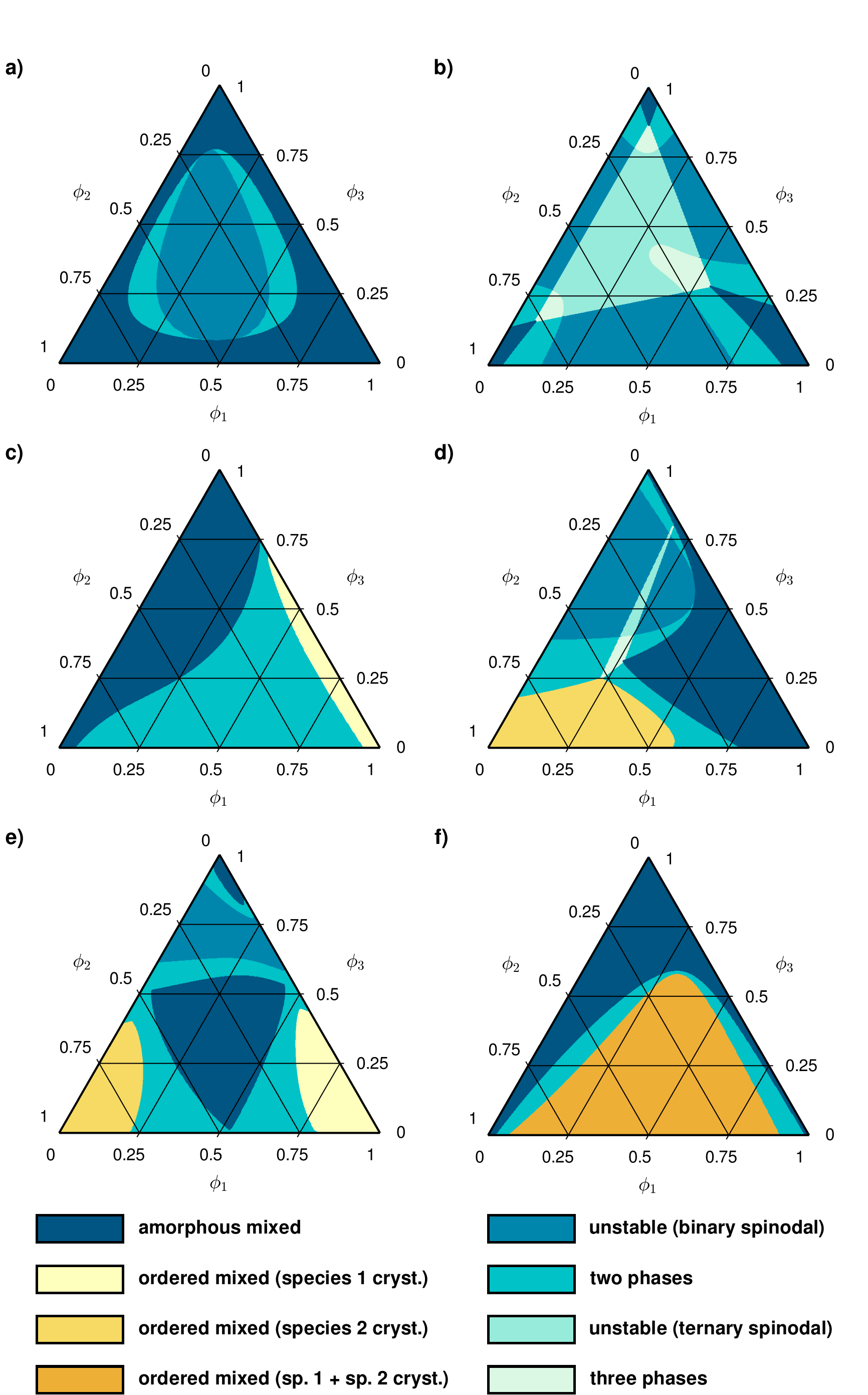}
    \caption{Phase diagrams of ternary mixtures exhibiting various distinct types of phase equilibria. All relevant parameters used for the calculation of the diagrams are provided in the SI (SI-D).}
    \label{fig:Ternary}
\end{figure}

Finally, phase diagrams computed for ternary systems are visualized in Fig.~\ref{fig:Ternary}. Fig.~\ref{fig:Ternary}-a depicts an amorphous mixture with a range of ternary compositions that are prone to phase separation despite all binary material combinations being fully miscible. Fig.~\ref{fig:Ternary}-b illustrates how amorphous miscibility gaps can overlap, leading either to binary or ternary phase equilibria with associated regions for binary and ternary spinodal decomposition. 

Fig.~\ref{fig:Ternary}-c and Fig.~\ref{fig:Ternary}-d then display selected crystallization configurations that involve only one crystalline component. In Fig.~\ref{fig:Ternary}-c, the crystallizing species experiences stronger repulsive interactions with the second material than with the third one. As a result, the two-phase region widens when the overall composition is close to that of a binary blend of components $1$ and $2$, and narrows progressively as it approaches the axis where the second species vanishes. In Fig.~\ref{fig:Ternary}-d, the amorphous-amorphous interactions between the second and the third constituent are sufficiently high to trigger the appearance of an amorphous demixing region. In this particular case, the interplay of the parameters causes a domain with a ternary phase equilibrium consisting of one crystalline and two amorphous phases. 

The mixtures represented in Fig.~\ref{fig:Ternary}-e and Fig.~\ref{fig:Ternary}-f include two species subject to crystallization. The three blend constituents in Fig.~\ref{fig:Ternary}-e are moderately incompatible both in the amorphous and/or in the crystalline state, so that all two-phase regions bridge from one diagram boundary to another. The central part of the diagram predicts a mixed amorphous phase due to the amorphous-amorphous interaction parameters being still low enough and the amorphous-crystalline and crystalline-amorphous ones being sufficiently high. Its area, however, tends to reduce when the former increase or the latter decrease. In comparison, the last figure (Fig.\ref{fig:Ternary}-f) presents a situation where the crystallizing components demonstrate relatively high compatibility in the ordered state. As already seen in Fig.\ref{fig:Binary2Cryst}-d for a binary blend, this can permit a composition range with a stable crystalline phase even above the melting temperatures of both pure materials.

All in all, these results demonstrate that a large variety of systems can be modelled with the derived free energy formulae. It may be mentioned that this showcase presentation of binary and ternary phase diagrams is by no means exhaustive and that many more shapes are available. Moreover, it has also to be stressed that all employed interaction parameters follow the linear form in $1/T$ with constant coefficients (Eq.~\ref{eq:Sec1_InteractionParamLinForm}). Allowing these to be more complex functions of temperature, composition and/or material properties is expected to extend the range of accessible blend behaviors even further.

\section{Conclusion}\label{Sec:Conclusion}

To summarize, this work presented a general free energy model describing the thermodynamics of mixing of crystalline multicomponent blends. By extending the mean-field approach commonly employed to calculate enthalpic mixing interactions between two amorphous species, the well-established Flory-Huggins theory was augmented to account for mixtures that involve any number of constituents, all of which being allowed to undergo a crystallization phase transition. Expressions for the chemical potentials of the blend components were also obtained from the derived free energy. In the limit of binary mixtures that exhibit perfectly pure crystalline phases, the chemical potentials were verified to consistently recover the melting point depression formula from the original theoretical framework. 

Notable features of the present model are the amorphous-crystalline, crystalline-amorphous, and crystalline-crystalline interaction parameters, which, in addition to the classical amorphous-amorphous one, determine miscibility properties between the mixed components depending on their respective state. A binary and ternary phase diagram showcase study demonstrated how these parameters impact phase separation phenomena that occur within blends. Depending on the interaction parameter values, interplay between miscibility gaps and melting point depressions, spinodal decomposition in the amorphous as well as in the crystalline state, and co-crystalline phase equilibria, can for instance be modelled.

An advantage of the current free energy formulation is to retain a relative simplicity, while being able to qualitatively represent various distinct and complex blend behaviors. It remains to be verified how accurately it can provide quantitative analyses for practical systems. Performing critical comparisons of model predictions against dedicated experimental measurements is therefore recommended for future investigations.

\section*{Author Contributions}

M. Siber: Conceptualization, Methodology, Investigation, Formal Analysis, Data Curation, Visualization, Writing - Original Draft, Writing - Review and Editing.\\O. J. J. Ronsin: Software, Supervision, Funding Acquisition, Writing - Review and Editing.\\J. Harting: Supervision, Project Administration, Funding Acquisition, Writing - Review and Editing. 

\section{Conflicts of interest}
There are no conflicts to declare.

\section{Acknowledgements}
The authors acknowledge financial support by the German Research Foundation (DFG, Project HA 4382/14-1), the European Commission (H2020 Program, Project 101008701/EMERGE), and the Helmholtz Association (SolarTAP Innovation Platform).

\newpage

\printbibliography[heading=bibintoc, title={References}]

@article{ameslon_taxonomy_2025,
  title = {Taxonomy of Amorphous Ternary Phase Diagrams: The Importance of Interaction Parameters},
  shorttitle = {Taxonomy of Amorphous Ternary Phase Diagrams},
  author = {Ameslon, Yasin and Liu, Hao and Harting, Jens and Ronsin, Olivier J. J. and Wodo, Olga},
  year = {2025},
  journal = {Physical Chemistry Chemical Physics},
  volume = {27},
  number = {19},
  pages = {9998--10010},
  issn = {1463-9076, 1463-9084},
  doi = {10.1039/D5CP00335K},
  urldate = {2025-05-22},
  langid = {english}
}

@article{aryanti_flory-huggins_2018,
  title = {Flory-{{Huggins Based Model}} to {{Determine Thermodynamic Property}} of {{Polymeric Membrane Solution}}},
  author = {Aryanti, P T P and Ariono, D and Hakim, A N and Wenten, I G},
  year = {2018},
  month = sep,
  journal = {Journal of Physics: Conference Series},
  volume = {1090},
  pages = {012074},
  issn = {1742-6588, 1742-6596},
  doi = {10.1088/1742-6596/1090/1/012074},
  urldate = {2021-01-04},
  langid = {english}
}

@article{baidakov_temperature_2013,
  title = {Temperature Dependence of the Crystal-Liquid Interfacial Free Energy and the Endpoint of the Melting Line},
  author = {Baidakov, Vladimir G. and Protsenko, Sergey P. and Tipeev, Azat O.},
  year = {2013},
  month = dec,
  journal = {The Journal of Chemical Physics},
  volume = {139},
  number = {22},
  pages = {224703},
  issn = {0021-9606},
  doi = {10.1063/1.4837695},
  urldate = {2025-07-31}
}

@article{barzin_theoretical_2007,
  title = {Theoretical Phase Diagram Calculation and Membrane Morphology Evaluation for Water/Solvent/Polyethersulfone Systems},
  author = {Barzin, J. and Sadatnia, B.},
  year = {2007},
  month = mar,
  journal = {Polymer},
  volume = {48},
  number = {6},
  pages = {1620--1631},
  issn = {0032-3861},
  doi = {10.1016/j.polymer.2007.01.049},
  urldate = {2025-08-15}
}

@article{blanks_thermodynamics_1964,
  title = {Thermodynamics of {{Polymer Solubility}} in {{Polar}} and {{Nonpolar Systems}}},
  author = {Blanks, R. F. and Prausnitz, J. M.},
  year = {1964},
  month = feb,
  journal = {Industrial \& Engineering Chemistry Fundamentals},
  volume = {3},
  number = {1},
  pages = {1--8},
  issn = {0196-4313, 1541-4833},
  doi = {10.1021/i160009a001},
  urldate = {2024-04-16},
  langid = {english}
}

@article{boom_equilibrium_1994,
  title = {Equilibrium {{Thermodynamics}} of a {{Quaternary Membrane-Forming System}} with {{Two Polymers}}. 1. {{Calculations}}},
  author = {Boom, R. M. and {van den Boomgaard}, {\relax Th}. and Smolders, C. A.},
  year = {1994},
  month = apr,
  journal = {Macromolecules},
  volume = {27},
  number = {8},
  pages = {2034--2040},
  publisher = {American Chemical Society},
  issn = {0024-9297},
  doi = {10.1021/ma00086a009},
  urldate = {2025-08-13}
}

@article{de_souza_exact_2024,
  title = {Exact Analytical Solution of the {{Flory}}--{{Huggins}} Model and Extensions to Multicomponent Systems},
  author = {{de Souza}, J. Pedro and Stone, Howard A.},
  year = {2024},
  month = jul,
  journal = {The Journal of Chemical Physics},
  volume = {161},
  number = {4},
  pages = {044902},
  issn = {0021-9606},
  doi = {10.1063/5.0215923},
  urldate = {2025-08-14}
}

@article{donnelly_probing_2015,
  title = {Probing the {{Effects}} of {{Experimental Conditions}} on the {{Character}} of {{Drug-Polymer Phase Diagrams Constructed Using Flory-Huggins Theory}}},
  author = {Donnelly, Conor and Tian, Yiwei and Potter, Catherine and Jones, David S. and Andrews, Gavin P.},
  year = {2015},
  month = jan,
  journal = {Pharmaceutical Research},
  volume = {32},
  number = {1},
  pages = {167--179},
  issn = {1573-904X},
  doi = {10.1007/s11095-014-1453-9},
  urldate = {2025-08-15},
  langid = {english}
}

@article{enders_interfacial_1994,
  title = {Interfacial Tension and Interaction Parameters},
  author = {Enders, S. and Huber, A. and Wolf, B. A.},
  year = {1994},
  month = jan,
  journal = {Polymer},
  volume = {35},
  number = {26},
  pages = {5743--5747},
  issn = {0032-3861},
  doi = {10.1016/S0032-3861(05)80050-3},
  urldate = {2025-07-31}
}

@article{favre_application_1996-1,
  title = {Application of {{Flory-Huggins}} Theory to Ternary Polymer-Solvents Equilibria: {{A}} Case Study},
  shorttitle = {Application of {{Flory-Huggins}} Theory to Ternary Polymer-Solvents Equilibria},
  author = {Favre, E. and Nguyen, Q. T. and Clement, R. and Neel, J.},
  year = {1996},
  month = mar,
  journal = {European Polymer Journal},
  volume = {32},
  number = {3},
  pages = {303--309},
  issn = {0014-3057},
  doi = {10.1016/0014-3057(95)00146-8},
  urldate = {2025-08-14}
}

@book{flory_principles_1953,
  title = {Principles of {{Polymer Chemistry}}},
  author = {Flory, Paul J.},
  year = {1953},
  publisher = {Cornell University Press},
  googlebooks = {CQ0EbEkT5R0C},
  isbn = {978-0-8014-0134-3},
  langid = {english}
}

@article{ghasemi_molecular_2021,
  title = {A Molecular Interaction--Diffusion Framework for Predicting Organic Solar Cell Stability},
  author = {Ghasemi, Masoud and Balar, Nrup and Peng, Zhengxing and Hu, Huawei and Qin, Yunpeng and Kim, Taesoo and Rech, Jeromy J. and Bidwell, Matthew and Mask, Walker and McCulloch, Iain and You, Wei and Amassian, Aram and Risko, Chad and O'Connor, Brendan T. and Ade, Harald},
  year = {2021},
  month = apr,
  journal = {Nature Materials},
  volume = {20},
  number = {4},
  pages = {525--532},
  publisher = {Nature Publishing Group},
  issn = {1476-4660},
  doi = {10.1038/s41563-020-00872-6},
  urldate = {2023-06-30},
  copyright = {2021 The Author(s), under exclusive licence to Springer Nature Limited},
  langid = {english}
}

@article{gottl_convex_2023,
  title = {Convex {{Envelope Method}} for Determining Liquid Multi-Phase Equilibria in Systems with Arbitrary Number of Components},
  author = {G{\"o}ttl, Quirin and Pirnay, Jonathan and Grimm, Dominik G. and Burger, Jakob},
  year = {2023},
  month = sep,
  journal = {Computers \& Chemical Engineering},
  volume = {177},
  pages = {108321},
  issn = {00981354},
  doi = {10.1016/j.compchemeng.2023.108321},
  urldate = {2024-09-27},
  langid = {english}
}

@article{granasy_phase-field_2019,
  title = {Phase-Field Modeling of Crystal Nucleation in Undercooled Liquids -- {{A}} Review},
  author = {Gr{\'a}n{\'a}sy, L{\'a}szl{\'o} and T{\'o}th, Gyula I. and Warren, James A. and Podmaniczky, Frigyes and Tegze, Gy{\"o}rgy and R{\'a}tkai, L{\'a}szl{\'o} and Pusztai, Tam{\'a}s},
  year = {2019},
  journal = {Progress in Materials Science},
  volume = {106},
  pages = {100569},
  issn = {0079-6425},
  doi = {10.1016/j.pmatsci.2019.05.002},
  urldate = {2021-12-13},
  langid = {english}
}

@article{hildebrand_history_1981,
  title = {A {{History}} of {{Solution Theory}}},
  author = {Hildebrand, J. H.},
  year = {1981},
  month = oct,
  journal = {Annual Review of Physical Chemistry},
  volume = {32},
  number = {1},
  pages = {1--24},
  issn = {0066-426X, 1545-1593},
  doi = {10.1146/annurev.pc.32.100181.000245},
  urldate = {2024-04-17},
  langid = {english}
}

@article{hildebrand_models_1953,
  title = {Models and Molecules. {{Seventh Spiers Memorial Lecture}}},
  author = {Hildebrand, Joel H.},
  year = {1953},
  month = jan,
  journal = {Discussions of the Faraday Society},
  volume = {15},
  number = {0},
  pages = {9--23},
  publisher = {The Royal Society of Chemistry},
  issn = {0366-9033},
  doi = {10.1039/DF9531500009},
  urldate = {2025-07-31},
  langid = {english}
}

@article{hildebrandt_predicting_1994-1,
  title = {Predicting Phase and Chemical Equilibrium Using the Convex Hull of the {{Gibbs}} Free Energy},
  author = {Hildebrandt, D. and Glasser, D.},
  year = {1994},
  month = jul,
  journal = {The Chemical Engineering Journal and the Biochemical Engineering Journal},
  volume = {54},
  number = {3},
  pages = {187--197},
  issn = {09230467},
  doi = {10.1016/0923-0467(94)00202-9},
  urldate = {2025-07-07},
  copyright = {https://www.elsevier.com/tdm/userlicense/1.0/},
  langid = {english}
}

@book{hillert_phase_2008,
  title = {Phase Equilibria, Phase Diagrams and Phase Transformations: Their Thermodynamic Basis},
  shorttitle = {Phase Equilibria, Phase Diagrams and Phase Transformations},
  author = {Hillert, Mats},
  year = {2008},
  publisher = {Cambridge University Press},
  address = {Cambridge, UK; New York},
  urldate = {2018-07-13},
  isbn = {978-1-139-12921-3 978-0-511-81278-1 978-0-511-50620-8},
  langid = {english}
}

@article{hohenberg_introduction_2015-1,
  title = {An Introduction to the {{Ginzburg}}--{{Landau}} Theory of Phase Transitions and Nonequilibrium Patterns},
  author = {Hohenberg, P. C. and Krekhov, A. P.},
  year = {2015},
  month = apr,
  journal = {Physics Reports},
  series = {An Introduction to the {{Ginzburg}}--{{Landau}} Theory of Phase Transitions and Nonequilibrium Patterns},
  volume = {572},
  pages = {1--42},
  issn = {0370-1573},
  doi = {10.1016/j.physrep.2015.01.001},
  urldate = {2025-07-29}
}

@article{horst_calculation_1995,
  title = {Calculation of Phase Diagrams Not Requiring the Derivatives of the {{Gibbs}} Energy Demonstrated for a Mixture of Two Homopolymers with the Corresponding Copolymer},
  author = {Horst, Roland},
  year = {1995},
  month = may,
  journal = {Macromolecular Theory and Simulations},
  volume = {4},
  number = {3},
  pages = {449--458},
  issn = {10221344, 15213919},
  doi = {10.1002/mats.1995.040040304},
  urldate = {2022-03-16},
  langid = {english}
}

@article{horst_calculation_1996,
  title = {Calculation of Phase Diagrams Not Requiring the Derivatives of the {{Gibbs}} Energy for Multinary Mixtures},
  author = {Horst, Roland},
  year = {1996},
  month = sep,
  journal = {Macromolecular Theory and Simulations},
  volume = {5},
  number = {5},
  pages = {789--800},
  issn = {10221344, 15213919},
  doi = {10.1002/mats.1996.040050501},
  urldate = {2022-03-16},
  langid = {english}
}

@article{horst_phase_1995-1,
  title = {Phase Diagrams Calculated for Quaternary Polymer Blends},
  author = {Horst, Roland and Wolf, B. A.},
  year = {1995},
  month = sep,
  journal = {The Journal of Chemical Physics},
  volume = {103},
  number = {9},
  pages = {3782--3787},
  issn = {0021-9606},
  doi = {10.1063/1.470708},
  urldate = {2025-08-13}
}

@article{hsu_thermodynamics_1974,
  title = {Thermodynamics of {{Polymer Compatibility}} in {{Ternary Systems}}},
  author = {Hsu, C. C. and Prausnitz, J. M.},
  year = {1974},
  month = may,
  journal = {Macromolecules},
  volume = {7},
  number = {3},
  pages = {320--324},
  publisher = {American Chemical Society},
  issn = {0024-9297},
  doi = {10.1021/ma60039a012},
  urldate = {2025-08-13}
}

@article{huggins_theory_1942,
  title = {Theory of {{Solutions}} of {{High Polymers1}}},
  author = {Huggins, Maurice L.},
  year = {1942},
  month = jul,
  journal = {Journal of the American Chemical Society},
  volume = {64},
  number = {7},
  pages = {1712--1719},
  publisher = {American Chemical Society},
  issn = {0002-7863},
  doi = {10.1021/ja01259a068},
  urldate = {2024-04-16}
}

@article{jian_temperature_2012,
  title = {Temperature Dependence of the Crystal--Melt Interfacial Energy of Metals},
  author = {Jian, Zengyun and Li, Na and Zhu, Man and Chen, Ji and Chang, Fange and Jie, Wanqi},
  year = {2012},
  month = may,
  journal = {Acta Materialia},
  volume = {60},
  number = {8},
  pages = {3590--3603},
  issn = {1359-6454},
  doi = {10.1016/j.actamat.2012.02.038},
  urldate = {2025-07-31}
}

@article{jung_polymerization_2010,
  title = {Polymerization of Methyl Methacrylate in the Presence of a Nonpolar Hydrocarbon Solvent. {{I}}. {{Construction}} of a Complete Ternary Phase Diagram through an {\emph{in Situ}} Polymerization},
  author = {Jung, Yunju and Luciani, Carla Vanesa and Han, Joong Jin and Choi, Kyu Yong},
  year = {2010},
  journal = {Journal of Applied Polymer Science},
  pages = {NA-NA},
  issn = {00218995, 10974628},
  doi = {10.1002/app.31924},
  urldate = {2019-09-11},
  langid = {english}
}

@article{kim_phase_2025,
  title = {Phase {{Behavior}} of {{Conjugated Polymer Solutions}} and {{Blends}}: {{The Flory}}--{{Huggins Lattice Model}}},
  shorttitle = {Phase {{Behavior}} of {{Conjugated Polymer Solutions}} and {{Blends}}},
  author = {Kim, Jung Yong},
  year = {2025},
  month = feb,
  journal = {The Journal of Physical Chemistry C},
  volume = {129},
  number = {7},
  pages = {3983--3992},
  publisher = {American Chemical Society},
  issn = {1932-7447},
  doi = {10.1021/acs.jpcc.5c00343},
  urldate = {2025-08-14}
}

@article{knopp_comparative_2015,
  title = {Comparative {{Study}} of {{Different Methods}} for the {{Prediction}} of {{Drug}}--{{Polymer Solubility}}},
  author = {Knopp, Matthias Manne and Tajber, Lidia and Tian, Yiwei and Olesen, Niels Erik and Jones, David S. and Kozyra, Agnieszka and L{\"o}bmann, Korbinian and Paluch, Krzysztof and Brennan, Claire Marie and Holm, Ren{\'e} and Healy, Anne Marie and Andrews, Gavin P. and Rades, Thomas},
  year = {2015},
  month = sep,
  journal = {Molecular Pharmaceutics},
  volume = {12},
  number = {9},
  pages = {3408--3419},
  publisher = {American Chemical Society},
  issn = {1543-8384},
  doi = {10.1021/acs.molpharmaceut.5b00423},
  urldate = {2025-08-15}
}

@article{konig_two-dimensional_2021,
  title = {Two-Dimensional {{Cahn}}--{{Hilliard}} Simulations for Coarsening Kinetics of Spinodal Decomposition in Binary Mixtures},
  author = {K{\"o}nig, Bj{\"o}rn and Ronsin, Olivier J. J. and Harting, Jens},
  year = {2021},
  journal = {Physical Chemistry Chemical Physics},
  volume = {23},
  number = {43},
  pages = {24823--24833},
  issn = {1463-9076, 1463-9084},
  doi = {10.1039/D1CP03229A},
  urldate = {2021-11-18},
  langid = {english}
}

@article{lauritzen_theory_1960,
  title = {Theory of {{Formation}} of {{Polymer Crystals}} with {{Folded Chains}} in {{Dilute Solution}}},
  author = {Lauritzen, John I. and Hoffman, John D.},
  year = {1960},
  journal = {Journal of Research of the National Bureau of Standards. Section A, Physics and Chemistry},
  volume = {64A},
  number = {1},
  pages = {73--102},
  issn = {0022-4332},
  doi = {10.6028/jres.064A.007},
  urldate = {2022-12-13},
  pmcid = {PMC5287029},
  pmid = {32196159}
}

@article{loo_composition_2019,
  title = {Composition {{Dependence}} of the {{Flory}}--{{Huggins Interaction Parameters}} of {{Block Copolymer Electrolytes}} and the {{Isotaksis Point}}},
  author = {Loo, Whitney S. and Sethi, Gurmukh K. and Teran, Alexander A. and Galluzzo, Michael D. and Maslyn, Jacqueline A. and Oh, Hee Jeung and Mongcopa, Katrina I. and Balsara, Nitash P.},
  year = {2019},
  month = aug,
  journal = {Macromolecules},
  volume = {52},
  number = {15},
  pages = {5590--5601},
  publisher = {American Chemical Society},
  issn = {0024-9297},
  doi = {10.1021/acs.macromol.9b00884},
  urldate = {2025-07-31}
}

@article{mao_phase_2019,
  title = {Phase Behavior and Morphology of Multicomponent Liquid Mixtures},
  author = {Mao, Sheng and Kuldinow, Derek and Haataja, Mikko P. and Ko{\v s}mrlj, Andrej},
  year = {2019},
  journal = {Soft Matter},
  volume = {15},
  number = {6},
  pages = {1297--1311},
  issn = {1744-683X, 1744-6848},
  doi = {10.1039/C8SM02045K},
  urldate = {2022-09-14},
  langid = {english}
}

@article{matkar_phase_2006,
  title = {Phase {{Diagrams}} of {{Binary Crystalline}}-{{Crystalline Polymer Blends}}},
  author = {Matkar, Rushikesh A. and Kyu, Thein},
  year = {2006},
  month = aug,
  journal = {The Journal of Physical Chemistry B},
  volume = {110},
  number = {32},
  pages = {16059--16065},
  publisher = {American Chemical Society},
  issn = {1520-6106},
  doi = {10.1021/jp062124p},
  urldate = {2022-01-14}
}

@article{matkar_role_2006,
  title = {Role of {{Crystal}}-{{Amorphous Interaction}} in {{Phase Equilibria}} of {{Crystal}}-{{Amorphous Polymer Blends}}},
  author = {Matkar, Rushikesh A. and Kyu, Thein},
  year = {2006},
  month = jun,
  journal = {The Journal of Physical Chemistry B},
  volume = {110},
  number = {25},
  pages = {12728--12732},
  publisher = {American Chemical Society},
  issn = {1520-6106},
  doi = {10.1021/jp061159m},
  urldate = {2022-01-14}
}

@article{nedoma_measurements_2008,
  title = {Measurements of the {{Composition}} and {{Molecular Weight Dependence}} of the {{Flory}}-{{Huggins Interaction Parameter}}},
  author = {Nedoma, Alisyn J. and Robertson, Megan L. and Wanakule, Nisita S. and Balsara, Nitash P.},
  year = {2008},
  month = aug,
  journal = {Macromolecules},
  volume = {41},
  number = {15},
  pages = {5773--5779},
  publisher = {American Chemical Society},
  issn = {0024-9297},
  doi = {10.1021/ma800698r},
  urldate = {2025-07-31}
}

@article{nishi_melting_1975,
  title = {Melting {{Point Depression}} and {{Kinetic Effects}} of {{Cooling}} on {{Crystallization}} in {{Poly}}(Vinylidene Fluoride)-{{Poly}}(Methyl Methacrylate) {{Mixtures}}},
  author = {Nishi, T. and Wang, T. T.},
  year = {1975},
  month = nov,
  journal = {Macromolecules},
  volume = {8},
  number = {6},
  pages = {909--915},
  issn = {0024-9297, 1520-5835},
  doi = {10.1021/ma60048a040},
  urldate = {2025-07-31},
  langid = {english}
}

@article{peng_materials_2023-1,
  title = {A Materials Physics Perspective on Structure--Processing--Function Relations in Blends of Organic Semiconductors},
  author = {Peng, Zhengxing and Stingelin, Natalie and Ade, Harald and Michels, Jasper J.},
  year = {2023},
  month = mar,
  journal = {Nature Reviews Materials},
  issn = {2058-8437},
  doi = {10.1038/s41578-023-00541-5},
  urldate = {2023-03-04},
  langid = {english}
}

@article{petri_composition-dependent_1995,
  title = {Composition-Dependent {{Flory-Huggins}} Parameters: Molecular Weight Influences at High Concentrations},
  shorttitle = {Composition-Dependent {{Flory-Huggins}} Parameters},
  author = {Petri, Hans-Michael and Wolf, B. A.},
  year = {1995},
  journal = {Macromolecular Chemistry and Physics},
  volume = {196},
  number = {7},
  pages = {2321--2333},
  issn = {1521-3935},
  doi = {10.1002/macp.1995.021960719},
  urldate = {2025-08-14},
  copyright = {{\copyright} 1995 H{\"u}thig \& Wepf Verlag, Zug},
  langid = {english}
}

@article{potter_investigation_2018,
  title = {Investigation of the {{Dependence}} of the {{Flory}}--{{Huggins Interaction Parameter}} on {{Temperature}} and {{Composition}} in a {{Drug}}--{{Polymer System}}},
  author = {Potter, Catherine B. and Davis, Mark T. and Albadarin, Ahmad B. and Walker, Gavin M.},
  year = {2018},
  month = nov,
  journal = {Molecular Pharmaceutics},
  volume = {15},
  number = {11},
  pages = {5327--5335},
  publisher = {American Chemical Society},
  issn = {1543-8384},
  doi = {10.1021/acs.molpharmaceut.8b00797},
  urldate = {2025-08-15}
}

@article{qian_analytical_2022-1,
  title = {Analytical {{Solution}} to the {{Flory}}--{{Huggins Model}}},
  author = {Qian, Daoyuan and Michaels, Thomas C. T. and Knowles, Tuomas P. J.},
  year = {2022},
  month = aug,
  journal = {The Journal of Physical Chemistry Letters},
  volume = {13},
  number = {33},
  pages = {7853--7860},
  publisher = {American Chemical Society},
  doi = {10.1021/acs.jpclett.2c01986},
  urldate = {2025-08-14}
}

@article{romay_thermodynamic_2021,
  title = {Thermodynamic {{Modeling}} and {{Validation}} of the {{Temperature Influence}} in {{Ternary Phase Polymer Systems}}},
  author = {Romay, Marta and Diban, Nazely and Urtiaga, Ane},
  year = {2021},
  month = jan,
  journal = {Polymers},
  volume = {13},
  number = {5},
  pages = {678},
  publisher = {Multidisciplinary Digital Publishing Institute},
  issn = {2073-4360},
  doi = {10.3390/polym13050678},
  urldate = {2025-08-13},
  copyright = {http://creativecommons.org/licenses/by/3.0/},
  langid = {english}
}

@article{ronsin_formation_2022,
  title = {Formation of {{Crystalline Bulk Heterojunctions}} in {{Organic Solar Cells}}: {{Insights}} from {{Phase-Field Simulations}}},
  shorttitle = {Formation of {{Crystalline Bulk Heterojunctions}} in {{Organic Solar Cells}}},
  author = {Ronsin, Olivier J. J. and Harting, Jens},
  year = {2022},
  journal = {ACS Applied Materials \& Interfaces},
  volume = {14},
  number = {44},
  pages = {49785-49800},
  issn = {1944-8244, 1944-8252},
  doi = {10.1021/acsami.2c14319},
  urldate = {2022-11-15},
  langid = {english}
}

@article{ronsin_phase-field_2022,
  title = {Phase-{{Field Simulations}} of the {{Morphology Formation}} in {{Evaporating Crystalline Multicomponent Films}}},
  author = {Ronsin, Olivier J. J. and Harting, Jens},
  year = {2022},
  journal = {Advanced Theory and Simulations},
  pages = {2200286},
  issn = {2513-0390},
  doi = {10.1002/adts.202200286},
  urldate = {2022-07-29},
  langid = {english}
}

@book{rubinstein_polymer_2003,
  title = {Polymer {{Physics}}},
  author = {Rubinstein, Michael and Colby, Ralph H.},
  year = {2003},
  month = jun,
  publisher = {Oxford University Press},
  address = {Oxford, New York},
  isbn = {978-0-19-852059-7}
}

@article{siber_crystalline_2023,
  title = {Crystalline Morphology Formation in Phase-Field Simulations of Binary Mixtures},
  author = {Siber, Maxime and Ronsin, Olivier J. J. and Harting, Jens},
  year = {2023},
  journal = {Journal of Materials Chemistry C},
  volume = {11},
  number = {45},
  pages = {15979--15999},
  publisher = {Royal Society of Chemistry},
  doi = {10.1039/D3TC03047D},
  urldate = {2023-11-23},
  langid = {english}
}

@article{takaki_phase-field_2014,
  title = {Phase-Field {{Modeling}} and {{Simulations}} of {{Dendrite Growth}}},
  author = {Takaki, Tomohiro},
  year = {2014},
  month = feb,
  journal = {ISIJ International},
  volume = {54},
  pages = {437--444},
  doi = {10.2355/isijinternational.54.437}
}

@article{tambasco_blend_2006,
  title = {Blend {{Miscibility}} and the {{Flory}}-{{Huggins Interaction Parameter}}:\, {{A Critical Examination}}},
  shorttitle = {Blend {{Miscibility}} and the {{Flory}}-{{Huggins Interaction Parameter}}},
  author = {Tambasco, Michael and Lipson, J. E. G. and Higgins, Julia S.},
  year = {2006},
  month = jul,
  journal = {Macromolecules},
  volume = {39},
  number = {14},
  pages = {4860--4868},
  publisher = {American Chemical Society},
  issn = {0024-9297},
  doi = {10.1021/ma060304r},
  urldate = {2025-08-14}
}

@article{thakore_analytical_2021,
  title = {Analytical and {{Computational Methods}} for the {{Determination}} of {{Drug-Polymer Solubility}} and {{Miscibility}}},
  author = {Thakore, Samarth D. and Akhtar, Junia and Jain, Ranjna and Paudel, Amrit and Bansal, Arvind K.},
  year = {2021},
  month = aug,
  journal = {Molecular Pharmaceutics},
  volume = {18},
  number = {8},
  pages = {2835--2866},
  issn = {1543-8384, 1543-8392},
  doi = {10.1021/acs.molpharmaceut.1c00141},
  urldate = {2025-08-15},
  copyright = {https://doi.org/10.15223/policy-029},
  langid = {english}
}

@article{thompson_approximation_1979,
  title = {On the Approximation of the Free Energy Change on Crystallization},
  author = {Thompson, Carl V. and Spaepen, Frans},
  year = {1979},
  month = dec,
  journal = {Acta Metallurgica},
  volume = {27},
  number = {12},
  pages = {1855--1859},
  issn = {0001-6160},
  doi = {10.1016/0001-6160(79)90076-2},
  urldate = {2025-07-31}
}

@article{turnbull_formation_1950,
  title = {Formation of {{Crystal Nuclei}} in {{Liquid Metals}}},
  author = {Turnbull, D.},
  year = {1950},
  month = oct,
  journal = {Journal of Applied Physics},
  volume = {21},
  number = {10},
  pages = {1022--1028},
  publisher = {American Institute of Physics},
  issn = {0021-8979},
  doi = {10.1063/1.1699435},
  urldate = {2022-02-28}
}

@article{venetsanos_mixing_2022-1,
  title = {Mixing {{Thermodynamics}} and {{Flory}}--{{Huggins Interaction Parameter}} of {{Polyethylene Oxide}}/{{Polyethylene Oligomeric Blends}} from {{Kirkwood}}--{{Buff Theory}} and {{Molecular Simulations}}},
  author = {Venetsanos, Fotis and Anogiannakis, Stefanos D. and Theodorou, Doros N.},
  year = {2022},
  month = jun,
  journal = {Macromolecules},
  volume = {55},
  number = {11},
  pages = {4852--4862},
  publisher = {American Chemical Society},
  issn = {0024-9297},
  doi = {10.1021/acs.macromol.2c00642},
  urldate = {2025-08-14}
}

@article{voskov_ternapi_2015,
  title = {{{TernAPI}} Program for the Calculation of Ternary Phase Diagrams with Isolated Miscibility Gaps by the Convex Hull Method},
  author = {Voskov, Alexey L and Dzuban, Alexander V and Maksimov, Alexey I},
  year = {2015},
  journal = {Fluid Phase Equilibria},
  pages = {9},
  langid = {english}
}

@article{vrij_equation_1968,
  title = {Equation for the Interfacial Tension between Demixed Polymer Solutions},
  author = {Vrij, A.},
  year = {1968},
  journal = {Journal of Polymer Science Part A-2: Polymer Physics},
  volume = {6},
  number = {11},
  pages = {1919--1932},
  issn = {1542-9377},
  doi = {10.1002/pol.1968.160061107},
  urldate = {2025-07-31},
  copyright = {Copyright {\copyright} 1968 John Wiley \& Sons, Inc.},
  langid = {english}
}

@article{willis_simple_2020-1,
  title = {Simple and {{Accurate Calibration}} of the {{Flory}}--{{Huggins Interaction Parameter}}},
  author = {Willis, James D. and Beardsley, Tom M. and Matsen, Mark W.},
  year = {2020},
  month = nov,
  journal = {Macromolecules},
  volume = {53},
  number = {22},
  pages = {9973--9982},
  publisher = {American Chemical Society},
  issn = {0024-9297},
  doi = {10.1021/acs.macromol.0c02115},
  urldate = {2025-08-15}
}

@article{xu_simultaneous_2014,
  title = {Simultaneous Determination of Three {{Flory}}--{{Huggins}} Interaction Parameters in Polymer/Solvent/Nonsolvent Systems by Viscosity and Cloud Point Measurements},
  author = {Xu, Li and Qiu, Feng},
  year = {2014},
  month = dec,
  journal = {Polymer},
  volume = {55},
  number = {26},
  pages = {6795--6802},
  issn = {0032-3861},
  doi = {10.1016/j.polymer.2014.10.045},
  urldate = {2025-08-13}
}

\end{document}